\documentclass[12pt]{article}
\usepackage{amsmath,mathtools,amsthm}
\usepackage{subfigure}

\begin{document}
\title{Bulk viscous cosmological model in Brans Dicke theory with new form of time varying deceleration parameter}
\author{G. P. Singh \footnote{gpsingh@mth.vnit.ac.in{GPS}}       \and
        Binaya K. Bishi\footnote{binaybc$@$gmail.com(BKB)}}

\date{ Department of Mathematics,
      Visvesvaraya National Institute of Technology,
       Nagpur-440010,INDIA\\\today}
\maketitle
\abstract{In this article we have presented  FRW cosmological model in the framework of Brans-Dicke theory.  This paper deals with a new proposed form of deceleration parameter and cosmological constant $\Lambda$. The effect of bulk viscosity is also studied in the presence of  modified Chaplygin gas  equation of state ($p=A\rho-\frac{B}{\rho^n}$). Further, we have discussed the  physical behaviours of the models.
}
\section{Introduction}
\label{intro}
It has been well established that alternative theories of gravitation played an important role in understanding the models of the Universe.  Since last few decades, researchers have shown more interest in alternative theories of gravitation especially scalar-tensor theories of gravity. The Brans-Dicke theory (BDT) of gravity is the one of the most successful alternative theory among all alternative theories of gravitation. This theory is consisting of a massless scalar field $\phi$ and a dimensionless constant $\omega$ describing the strength of the coupling between $\phi$   and the matter \cite{BransDicke61}. In the BDT, gravitational constant $G$ is treated as the reciprocal of a massless scalar field $\phi$, where $\phi$ is expected to satisfy a scalar wave equations and it's source is all matter in the Universe.
\par
In a pioneering work, both research contributions by Mathiazhagan \& Johri\cite{MathiazhaganJohri84} and later La \& Steinhardt \cite{LaSteinhardt89} showed that the idea of inflationary expansion with a first order phase transition can be made to work more satisfactorily if one considers the BDT  in place of general relativity. The interesting consequence of  BD scalar field is that the modified field equations would express the scale factor $R(t)$ as a power function of time and not as an exponential function, so that one attains the so-called ``graceful exit" from the inflationary vacuum phase through a first order phase transition. Hyperextend inflation \cite{SteinhardtAccetta90} generalize the results of extended inflation in BDT and solves the graceful exit problem in a natural way, without recourse to any fine tuning as required in relativistic models. Romero \& Barros \cite{RomeroBarros93} discussed about the limit of the Brans-Dicke theory of gravity when $\omega\rightarrow\infty$  and shown by examples that, in this limit it is not always true that BDT reduces to general relativity. From the literature, it is known that the result of BDT is close to Einstein theory of general relativity for large value of the coupling parameter $(\omega\geq500)$ \cite{Will81,Faraoni04}. A more recent bound on the Brans-Dicke parameter $\omega$ is $\omega >3300$ \cite{Faraoni04}. A number of researchers \cite{JohriDesikan94,BarrowMagueijo99,SinghBeesham99,SenBanerjee00,BarrosRomero01,ChakrabortyGhosh03,ReddyLakshmi14,ShamirBhatti12} have discussed various aspects of expanding cosmological models in BDT.
\par
Cosmological observations \cite{Knop03,Bennet03} and various related research clearly indicate that, the constituent of the  present Universe is dominated by dark energy, which constitutes about three fourths of the whole matter of our Universe.  There are several candidates for dark energy like quintessence, phantom, quintom, holographic dark energy, K-essence, Chaplygin  gas and cosmological constant. Among all the dark energy candidates, cosmological constant is the more favoured. It provides enough negative pressure to account the acceleration and contribute an energy density of same order of magnitude than the energy density of the matter \cite{Vishwakarma02}. The discrepancy of observed value and theoretical value of cosmological constant is usually referred as cosmological constant problem in literature. This problem is the puzzling problem in standard cosmology.  The cosmological constant bears a dynamical decaying character so that it might be large at early epoch and approaching to a small value at the present epoch.
\par
The effect of cosmological constant has been discussed in the literature in the context of general relativity and its alternative theories. Singh \& Singh \cite{SinghSingh84} presented a cosmological model in BDT by considering cosmological constant as a function of scalar field $\phi$. Exact cosmological solutions in BDT with uniform cosmological “constant”  has been studied by Pimentel \cite{Pimentel85}. A class of flat FRW cosmological models with cosmological “constant” in BDT have also been obtained by Ahmadi \& Riazi \cite{AhmadiRiazi95}. The age of the Universe from a view point of the nucleosynthesis with $\Lambda$ term in BDT was investigated by Etoh et al. \cite{Etoh97}. Azad \& Islam. \cite{AzadIslam03} extended the idea of Singh \& Singh \cite{SinghSingh84} to study cosmological constant in Bianchi type I modified Brans-Dicke cosmology. Qiang \cite{Qiang05} discussed cosmic acceleration in five dimensional BDT using interacting Higgs and Brans-Dicke fields. Smolyakov \cite{Smolyakov07} investigated a model which provides the necessary value of effective cosmological “constant” at the classical level. Recently, embedding general relativity with varying cosmological term in five dimensional BDT of gravity in vacuum has been discussed by Reyes \& Aguilar \cite{ReyesAguilar09}. Singh et al. \cite{Singh13} have studied the dynamic cosmological constant in BDT.
\par
On the other side, it is known from the literature that for early evolution of the Universe, bulk viscosity is supposed to play a very important role. The presence of viscosity in the fluid explore many dynamics of the homogeneous cosmological models. The bulk viscosity coefficient determine the magnitude of the viscous stress relative to the expansion. Recently Saadat \& Pourhassan \cite{SaadatPourhassan13} investigated the FRW bulk viscous cosmology with modified cosmic Chaplygin gas. Many researchers also have shown interest in FRW bulk viscous cosmological models in different contexts (see \cite{SaadatPourhassan13} and references there in).
\par
Motivated by the above studies, here we have discussed the variable cosmological constant $\Lambda$ for FRW metric in the context of BDT with a special form of deceleration parameter.
\section{Field equations}
The field equation of Brans-Dicke theory in presence of cosmological constant may be written as
\begin{eqnarray}
\label{eqn1}
G_{ij}-\Lambda g_{ij}+\frac{\omega}{\phi^2}\left[\phi_{;i}\phi_{;j}-\frac{1}{2}g_{ij}\phi_{;k}\phi^{;k}\right]
+\frac{1}{\phi}\left[\phi_{;i;j}-g_{ij}\qed\phi\right]=\frac{8\pi}{\phi}T_{ij},
\end{eqnarray}
\begin{equation}
\label{eqn2}
\qed\phi=\phi_{;i}^{;i}=\frac{8\pi}{2\omega+3}T_{,i}^i
\end{equation}
where\,$\phi$ is the scalar field. The energy-momentum tensor $T_{ij}$ of the cosmic fluid in the presence of bulk viscosity may be
be defined as
\begin{equation}
\label{eqn3}
T_{ij}=(\rho+p+\Pi)u_iu_j-(p+\Pi)g_{ij}
\end{equation}
Let us consider a homogeneous and isotropic Universe represented by FRW spacetime
metric as
\begin{equation}
\label{eqn4}
ds^2=dt^2-R^2(t)\left[\frac{dr^2}{1-kr^2}+r^2\left(d\theta^2+\sin^2\theta d\phi^2\right)\right]
\end{equation}
where $k$ $(=1,0,-1)$ is the curvature parameter, which represents closed, flat and open model of the Universe and $R(t)$ is the scale factor.\\
The FRW metric \eqref{eqn4} and energy-momentum tensor \eqref{eqn3} along with Brans-Dicke
field equations yield the following equations
\begin{equation}
\label{eqn5}
3\left(\frac{\dot{R}}{R}\right)^2+3\frac{\dot{R}\dot{\phi}}{R\phi}-\frac{\omega}{2}\left(\frac{\dot{\phi}}{\phi}\right)^2+3\frac{k}{R^2}=\frac{8\pi}{\phi}\rho+\Lambda
\end{equation}
\begin{equation}
\label{eqn6}
2\frac{\ddot{R}}{R}+\left(\frac{\dot{R}}{R}\right)^2+\frac{\ddot{\phi}}{\phi}+\frac{\omega}{2}\left(\frac{\dot{\phi}}{\phi}\right)^2+2\frac{\dot{R}\dot{\phi}}{R\phi}+\frac{k}{R^2}=\frac{-8\pi}{\phi}(p+\Pi)+\Lambda
\end{equation}
\begin{equation}
\label{eqn7}
\frac{\ddot{\phi}}{\phi}+3\frac{\dot{R}\dot{\phi}}{R\phi}=\frac{8\pi}{\phi}\frac{\rho-3p-3\Pi}{3+2\omega}+\frac{2\Lambda}{3+2\omega}
\end{equation}
\section{Solution of the field equations}
In order to find exact solutions of basic field equations \eqref{eqn5}-\eqref{eqn7}, one must ensure that set of equations should be closed. Thus, two more physically reasonable relations are required amongst the variables.\\First we consider a  well accepted power law relation
between scale factor\,$R(t)$\, and scalar field\,$\phi$\, of the form \cite{Singh13}
\begin{equation}
\label{eqn8}
\phi=\phi_0R^{\alpha_1}
\end{equation}
and as it has been well established that the expansion of present Universe is accelerating. In order to study a cosmological model with early deceleration and late time  acceleration, we have proposed deceleration parameter of the form
\begin{equation}
\label{eqn9}
q=\frac{\alpha_2+\alpha_3 t}{1+t}
\end{equation}
as the second physically plausible relation. Where $\alpha_2,\alpha_3\in \bf{R}$. The considered form of deceleration parameter is motivated by the bilinear form of deceleration parameter \cite{MishraChand16}. Deceleration parameter is useful to classify the models of the Universe. From literature we know that deceleration parameter is a constant quantity or it depends on time. In the case when rate of expansion never change and $\dot{R}$ is constant, the scaling factor is proportional to time, which leads to zero deceleration. In case when $H$ is constant, the deceleration parameter ($q$) is also constant (-1). In de-Sitter and steady state Universe such cases arises. Now we will classify the Cosmological models on the basis of time dependence on Hubble parameter and deceleration parameter as follows \cite{Bolotin15}.
\begin{enumerate}
  \item $H>0$,\; $q>0$: expanding and decelerating
  \item $H > 0$,\; $q < 0$: expanding and accelerating
  \item $H < 0$,\;$q > 0$: contracting and decelerating
  \item $H < 0$,\; $q < 0$: contracting and accelerating
  \item $H > 0$,\; $q = 0$: expanding, zero deceleration / constant expansion
  \item $H < 0$,\; $q = 0$: contracting, zero deceleration
  \item $H = 0$,\; $q = 0$: static
\end{enumerate}
From the above classification, 1,2 and 5 are possible cases as in the present scenario our Universe is expanding. Again also we have found the following type of expansion exhibit by our Universe.
\begin{enumerate}
  \item $q<-1$: super exponential expansion
  \item $-1\leq q < 0$: exponential expansion (for $q=-1$ known as de-Sitter expansion)
  \item $q=0$: expansion with constant rate
  \item $-1<q <1$: accelerating power expansion
  \item $q>0$: decelerating expansion
\end{enumerate}
We consider third physically plausible relation as the modified Chaplygin gas equation of state as follows\cite{Debnath11,Samanta14}
\begin{equation}
\label{eqn10}
p=A\rho-\frac{B}{\rho^n}
\end{equation}
where $A>0$, $B>0$ are constants and $0\leq n\leq 1$.\\
The set of field equations \eqref{eqn5}-\eqref{eqn7} with the help of \eqref{eqn8} may be written as
\begin{equation}
\label{eqn11}
\left(\frac{6+6\alpha_1-\omega\alpha_1^2}{2}\right)\left(\frac{\dot{R}}{R}\right)^2+\frac{3k}{R^2}=\frac{8\pi}{\phi_0R^{\alpha_1}}\rho+\Lambda
\end{equation}
\begin{equation}
\begin{split}
\label{eqn12}
(2+\alpha_1)\frac{\ddot{R}}{R}+\left(\frac{2+2\alpha_1+2\alpha_1^2+\omega\alpha_1^2}{2}\right)\left(\frac{\dot{R}}{R}\right)^2+\frac{k}{R^2}=\frac{-8\pi}{\phi_0R^{\alpha_1}}(p+\Pi)+\Lambda
\end{split}
\end{equation}
\begin{equation}
\label{eqn13}
\begin{split}
\left[\alpha_1\frac{\ddot{R}}{R}+\alpha_1(\alpha_1+2)\left(\frac{\dot{R}}{R}\right)^2\right](3+2\omega)=\frac{8\pi}{\phi_0R^{\alpha_1}}(\rho-3p-3\Pi)+2\Lambda
\end{split}
\end{equation}
Equations \eqref{eqn11},\eqref{eqn12} and \eqref{eqn13}, leads us to
\begin{equation}
\label{eqn14}
2(3-\omega\alpha_1)\frac{\ddot{R}}{R}+(6-4\omega\alpha_1-\omega\alpha_1^2)\left(\frac{\dot{R}}{R}\right)^2+\frac{6k}{R^2}=2\Lambda
\end{equation}
This equation is useful for obtaining the various cosmological solutions.
\par
Now our problem is to evaluate the $R(t)$, which is obtained from the relation
\begin{equation}\label{eqn15}
-\frac{\dot{H}}{H^2}=1+q.
\end{equation}
With the help of equation \eqref{eqn9} and integrating \eqref{eqn15}, we obtained
\begin{equation}\label{eqn16}
H=\frac{1}{(1+\alpha_3)t+(\alpha_2-\alpha_3)ln(1+t)+c_1},
\end{equation}
where $c_1$ is a constant of integration. The condition $H\rightarrow \infty$ when $t\rightarrow 0$ yields $c_1=0$. Thus, equation \eqref{eqn16} takes the form
\begin{equation}\label{eqn17}
H=\frac{1}{(1+\alpha_3)t+(\alpha_2-\alpha_3)ln(1+t)}
\end{equation}
Equation \eqref{eqn17} is expressed as
\begin{eqnarray*}
 H&=& \frac{1}{(1+\alpha_3 )t+(\alpha_2-\alpha_3)\left[t-\frac{t^2}{2}+\frac{t^3}{3}-\frac{t^4}{4}+\frac{t^5}{5}-\frac{t^6}{6}+\frac{t^7}{7}-\cdot\cdot\cdot\cdot\cdot\cdot\right]} \\
 {} &=& \frac{1}{(1+\alpha_2 )t+(\alpha_2-\alpha_3)\left[-\frac{t^2}{2}+\frac{t^3}{3}-\frac{t^4}{4}+\frac{t^5}{5}-\frac{t^6}{6}+\frac{t^7}{7}-\cdot\cdot\cdot\cdot\cdot\cdot\right]} \\
  {} &=& \frac{1}{(1+\alpha_2 )t}\bigg\{1-\frac{(\alpha_2-\alpha_3)}{1+\alpha_2}\left[\frac{t}{2}-\frac{t^2}{3}+\frac{t^3}{4}-\frac{t^4}{5}+\frac{t^5}{6}-\frac{t^6}{7}+\cdot\cdot\cdot\cdot\cdot\cdot\right]\bigg\}^{-1}
\end{eqnarray*}
\begin{eqnarray*}
  {} &=&  \frac{1}{(1+\alpha_2 )t}\bigg[1+\left(\frac{\alpha_2-\alpha_3}{1+\alpha_2}\right)\left(\frac{t}{2}-\frac{t^2}{3}+\frac{t^3}{4}-\frac{t^4}{5}+\frac{t^5}{6}-\frac{t^6}{7}+\cdot\cdot\cdot\cdot\cdot\cdot\right)\\ {} &+& \left(\frac{\alpha_2-\alpha_3}{1+\alpha_2}\right)^2\left(\frac{t}{2}-\frac{t^2}{3}+\frac{t^3}{4}-\frac{t^4}{5}+\frac{t^5}{6}-\frac{t^6}{7}+\cdot\cdot\cdot\cdot\cdot\cdot\right)^2 \\
   {} &+& \left(\frac{\alpha_2-\alpha_3}{1+\alpha_2}\right)^3\left(\frac{t}{2}-\frac{t^2}{3}+\frac{t^3}{4}-\frac{t^4}{5}+\frac{t^5}{6}-\frac{t^6}{7}+\cdot\cdot\cdot\cdot\cdot\cdot\right)^3 \\
  {} &+& \left(\frac{\alpha_2-\alpha_3}{1+\alpha_2}\right)^4\left(\frac{t}{2}-\frac{t^2}{3}+\frac{t^3}{4}-\frac{t^4}{5}+\frac{t^5}{6}-\frac{t^6}{7}+\cdot\cdot\cdot\cdot\cdot\cdot\right)^4 +\cdot\cdot\cdot\cdot\cdot\cdot\bigg] \\
\end{eqnarray*}
Simplifying the above expression we obtained
\begin{equation}\label{eqn18}
H=\frac{1}{(1+\alpha_2 )t}+k_0+k_1t+k_2t^2+k_3t^3+k_4t^4+O(t^5),
\end{equation}
where
\begin{eqnarray*}
  k_0 &=& \frac{\alpha_2-\alpha_3}{2(1+\alpha_2)^2} \\
   k_1 &=&\frac{1}{1+\alpha_2}\left[\frac{(\alpha_2-\alpha_3)^2}{4(1+\alpha_2)^2}-\frac{(\alpha_2-\alpha_3)}{3(1+\alpha_2)} \right]\\
  k_2&=&\frac{1}{1+\alpha_2}\left[\frac{\alpha_2-\alpha_3}{4(1+\alpha_2)}-\frac{(\alpha_2-\alpha_3)^2}{3(1+\alpha_2)^2}+\frac{(\alpha_2-\alpha_3)^3}{8(1+\alpha_2)^3}\right] \\
  k_3&=&\frac{1}{1+\alpha_2}\left[-\frac{\alpha_2-\alpha_3}{5(1+\alpha_2)}+\frac{13(\alpha_2-\alpha_3)^2}{36(1+\alpha_2)^2}-\frac{(\alpha_2-\alpha_3)^3}{4(1+\alpha_2)^3}+\frac{(\alpha_2-\alpha_3)^4}{16(1+\alpha_2)^4}\right] \\
  k_4&=&\frac{1}{1+\alpha_2}\left[\frac{\alpha_2-\alpha_3}{6(1+\alpha_2)}-\frac{11(\alpha_2-\alpha_3)^2}{30(1+\alpha_2)^2}+\frac{17(\alpha_2-\alpha_3)^3}{48(1+\alpha_2)^3}-\frac{(\alpha_2-\alpha_3)^4}{6(1+\alpha_2)^4}\right] \end{eqnarray*}
Integration of \eqref{eqn18} leads us to
\begin{equation}\label{eqn19}
R=c_2t^{\frac{1}{1+\alpha_2}}e^{T_1(t)},
\end{equation}
where $T_1(t)=k_0t+k_1\frac{t^2}{2}+k_2\frac{t^3}{3}+k_3\frac{t^4}{4}+k_4\frac{t^5}{5}+O(t^6).$ The solutions of the field equation \eqref{eqn11}-\eqref{eqn13} is expressed as follows:
The energy density $\rho$ is obtained as
\begin{equation}\label{eqn20}
\rho=\frac{k_5t^{\frac{\alpha_1}{1+\alpha_2}}e^{\alpha_1T_1(t)}(\rho_1+\rho_2t)}{(1+t)\left[(1+\alpha_3)t+(\alpha_2-\alpha_3)ln(1+t)\right]^2},
\end{equation}
where $k_5=\frac{\phi_0c_2^{\alpha_1}}{8\pi}$,\;$\rho_1=(3+2w-w\alpha_2)\alpha_1+3\alpha_2$,\;$\rho_2=(3+2w-w\alpha_3)\alpha_1+3\alpha_3$.\\
The pressure $p$ is given as
\begin{equation}\label{eqn21}
p=\frac{Ak_5^{n+1}t^{\frac{(n+1)\alpha_1}{1+\alpha_2}}e^{(n+1)\alpha_1T_1(t)}(\rho_1+\rho_2t)^{n+1}-B(1+t)^{n+1}\left[(1+\alpha_3)t+(\alpha_2-\alpha_3)ln(1+t)\right]^{2n+2}}{k_5^{n}t^{\frac{n\alpha_1}{1+\alpha_2}}e^{n\alpha_1T_1(t)}(\rho_1+\rho_2t)^{n}(1+t)\left[(1+\alpha_3)t+(\alpha_2-\alpha_3)ln(1+t)\right]^2}
\end{equation}
The bulk viscous stress $\Pi$ is expressed as
\begin{equation}\label{eqn22}
\Pi=\frac{\splitfrac{k_5^{n+1}t^{\frac{(n+1)\alpha_1}{1+\alpha_2}}e^{(n+1)\alpha_1T_1(t)}\left[\frac{\rho_1+\rho_2t}{(1+t)\left[(1+\alpha_3)t+(\alpha_2-\alpha_3)ln(1+t)\right]^2}\right]^n\times}{\left[\frac{\Pi_1+\Pi_2t}{(1+t)\left[(1+\alpha_3)t+(\alpha_2-\alpha_3)ln(1+t)\right]^2}+\frac{2k}{c_2^2t^{\frac{2}{1+\alpha_2}}e^{2T_1(t)}}\right]+B}}{k_5^{n}t^{\frac{n\alpha_1}{1+\alpha_2}}e^{n\alpha_1T_1(t)}\left[\frac{\rho_1+\rho_2t}{(1+t)\left[(1+\alpha_3)t+(\alpha_2-\alpha_3)ln(1+t)\right]^2}\right]^n},
\end{equation}
where $\Pi_1=-(1+w)\alpha_1^2-(1+3A)\alpha_2+[(\alpha_2-2)(1+A)w-1-3A+\alpha_2]\alpha_1+2$ and
$\Pi_2=-(1+w)\alpha_1^2-(1+3A)\alpha_3+[(\alpha_2-2)(1+A)w-1-3A+\alpha_3]\alpha_1+2$.\\
The cosmological constant $\Lambda$ is expressed as
\begin{equation}\label{eqn23}
\Lambda=\frac{\Lambda_1+\Lambda_2t}{(1+t)\left[(1+\alpha_3)t+(\alpha_2-\alpha_3)ln(1+t)\right]^2}+\frac{3k}{c_2^2t^{\frac{2}{1+\alpha_2}}e^{2T_1(t)}},
\end{equation}
where $\Lambda_1=-0.5w\alpha_1^2+w(\alpha_2-2)\alpha_1-3(\alpha_2-1)$ and $\Lambda_2=-0.5w\alpha_1^2+w(\alpha_3-2)\alpha_1-3(\alpha_3-1)$.
\begin{table}[ht!]
\centering
\begin{center}
\begin{tabular}{ |c|c|c|c|c| }
\hline
S. No.&Possible value of $\alpha_2$ and $\alpha_3$&\begin{tabular}{@{}c@{}}Form of deceleration\\ parameter $q$ \end{tabular}& Behaviour of Cosmological model\\
\hline
1&\begin{tabular}{@{}c@{}}$\alpha_2=0$ \\ $\alpha_3>0$\end{tabular}&$q=\frac{\alpha_3 t}{1+t}$&Decelerating\\
\hline
2 &\begin{tabular}{@{}c@{}}$\alpha_2=0$ \\ $\alpha_3<0$ $(\alpha_3=-\alpha_4$,$\alpha_4>0$)\end{tabular}&$q=-\frac{\alpha_4 t}{1+t}$&Accelerating\\
\hline
3&\begin{tabular}{@{}c@{}}$\alpha_2=0$ \\ $\alpha_3=0$\end{tabular}&$q=0$&\begin{tabular}{@{}c@{}}Expansion with\\ constant rate\end{tabular}\\
\hline
4&\begin{tabular}{@{}c@{}}$\alpha_2>0$ \\ $\alpha_3>0$\end{tabular}&$q=\frac{\alpha_2+\alpha_3 t}{1+t}$&Decelerating\\
\hline
5&\begin{tabular}{@{}c@{}}$\alpha_2>0$ \\ $\alpha_3<0$\end{tabular}&$q=\frac{\alpha_2-\alpha_4 t}{1+t}$&\begin{tabular}{@{}c@{}}Phase trasition from\\ decelerating to accelerating\end{tabular}\\
\hline
6&\begin{tabular}{@{}c@{}}$\alpha_2>0$ \\ $\alpha_3=0$\end{tabular}&$q=\frac{\alpha_2}{1+t}$&Decelerating\\
\hline
7&\begin{tabular}{@{}c@{}}$\alpha_2<0 (\alpha_2=-\alpha_5,\alpha_5>0)$ \\ $\alpha_3>0$\end{tabular}&$q=\frac{-\alpha_5+\alpha_3 t}{1+t}$&\begin{tabular}{@{}c@{}}Phase trasition from\\ accelerating to decelerating \end{tabular}\\
\hline
8&\begin{tabular}{@{}c@{}}$\alpha_2<0$ \\ $\alpha_3<0$\end{tabular}&$q=\frac{-\alpha_5-\alpha_4 t}{1+t}$& Accelerating \\
\hline
9&\begin{tabular}{@{}c@{}}$\alpha_2<0$ \\ $\alpha_3=0$\end{tabular}&$q=-\frac{\alpha_5}{1+t}$& Accelerating\\
\hline
\end{tabular}
\end{center}
\caption{Different forms of deceleration parameter depending on the parameters $\alpha_2$ \& $\alpha_3$ and behaviour of cosmological models according to the deceleration parameter $q$.}
\label{Tab1}
\end{table}
\par
 Now, let us start with our proposed form of deceleration parameter $q$. The different form of deceleration parameter is evolved as a result of considered value of $\alpha_2$ and $\alpha_3$, which is expressed in Table \ref{Tab1}. We know that in present scenario our Universe is accelerating. Thus serial numbers 2, 5, 8 and 9 of Table \ref{Tab1} exhibits accelerating model. Now we will discuss about the deceleration parameter in serial numbers 2, 5, 8 and 9 of Table \ref{Tab1}. For the choice of $\alpha_2=\alpha=\alpha_4$, the deceleration parameter $q$ in serial number 2 and 5 of Table \ref{Tab1} reduces to $q=-\frac{\alpha t}{1+t}$ and $q=\frac{-\alpha(1+ t)}{1+t}$ respectively, which is discussed by \cite{MishraChand16}. They called this deceleration parameter as Bilinear variable deceleration parameter. We will discuss the case  where $\alpha_2\neq\alpha \neq\alpha_4$ of serial number 5 of Table \ref{Tab1} and also serial number 8 and 9 of Table.\ref{Tab1}. According to the serial number 5, 8 and 9 of Table \ref{Tab1} we have three different models, which are discussed below.
  \subsection{Model-I}
 The deceleration parameter $q$ in \eqref{eqn9} for $\alpha_2>0$ and $\alpha_3<0$ takes the form
 \begin{equation}\label{eqn24}
q=\frac{\alpha_2-\alpha_4 t}{1+t},\;\;\alpha_3=-\alpha_4,\;\alpha_4>0
 \end{equation}
Here we noticed that, $q>0$ for $0<t<\frac{\alpha_2}{\alpha_4}$ and $q<0$ for $t>\frac{\alpha_2}{\alpha_4}$, which means that our Universe is decelerating and accelerating in the provided ranges respectively. Thus our Universe undergoes a phase transition from decelerating to accelerating phase.
\par
For model I, the physical parameters are obtained as follows:\\
The Hubble parameter in \eqref{eqn17} takes the form
\begin{equation}\label{eqn25}
H=\frac{1}{(1-\alpha_4)t+(\alpha_2+\alpha_4)ln(1+t)}
\end{equation}
The scale factor $R(t)$ in \eqref{eqn19} is expressed as
\begin{equation}\label{eqn26}
R=c_2t^{\frac{1}{1+\alpha_2}}e^{T_1(t)},
\end{equation}
where $T_1(t)=k_0t+k_1\frac{t^2}{2}+k_2\frac{t^3}{3}+k_3\frac{t^4}{4}+k_4\frac{t^5}{5}+O(t^6).$ and
\begin{eqnarray*}
  k_0 &=& \frac{\alpha_2+\alpha_4}{2(1+\alpha_2)^2} \\
   k_1 &=&\frac{1}{1+\alpha_2}\left[\frac{(\alpha_2+\alpha_4)^2}{4(1+\alpha_2)^2}-\frac{(\alpha_2+\alpha_4)}{3(1+\alpha_2)} \right]\\
  k_2&=&\frac{1}{1+\alpha_2}\left[\frac{\alpha_2+\alpha_4}{4(1+\alpha_2)}-\frac{(\alpha_2+\alpha_4)^2}{3(1+\alpha_2)^2}+\frac{(\alpha_2+\alpha_4)^3}{8(1+\alpha_2)^3}\right] \\
  k_3&=&\frac{1}{1+\alpha_2}\left[-\frac{\alpha_2+\alpha_4}{5(1+\alpha_2)}+\frac{13(\alpha_2+\alpha_4)^2}{36(1+\alpha_2)^2}-\frac{(\alpha_2+\alpha_4)^3}{4(1+\alpha_2)^3}+\frac{(\alpha_2+\alpha_4)^4}{16(1+\alpha_2)^4}\right] \\
  k_4&=&\frac{1}{1+\alpha_2}\left[\frac{\alpha_2+\alpha_4}{6(1+\alpha_2)}-\frac{11(\alpha_2+\alpha_4)^2}{30(1+\alpha_2)^2}+\frac{17(\alpha_2+\alpha_4)^3}{48(1+\alpha_2)^3}-\frac{(\alpha_2+\alpha_4)^4}{6(1+\alpha_2)^4}\right] \end{eqnarray*}
The FRW space-time metric in \eqref{eqn4} takes the form
\begin{equation*}
ds^2=dt^2-c_2^2t^{\frac{2}{1+\alpha_2}}e^{2T_1(t)}\left[\frac{dr^2}{1-kr^2}+r^2\left(d\theta^2+\sin^2\theta d\phi^2\right)\right]
\end{equation*}
with the above mentation $k_i$,$(i=0,1,2,3,4)$.
The energy density $(\rho)$, pressure $(p)$, bulk viscous stress $(\Pi)$ and cosmological constant $(\Lambda)$in \eqref{eqn20}, \eqref{eqn21}, \eqref{eqn22} and \eqref{eqn23} are expressed as
\begin{equation}\label{eqn27}
\rho=\frac{k_5t^{\frac{\alpha_1}{1+\alpha_2}}e^{\alpha_1T_1(t)}(\rho_1+\rho_2t)}{(1+t)\left[(1-\alpha_4)t+(\alpha_2+\alpha_4)ln(1+t)\right]^2},
\end{equation}
where $k_5=\frac{\phi_0c_2^{\alpha_1}}{8\pi}$,\;$\rho_1=(3+2w-w\alpha_2)\alpha_1+3\alpha_2$,\;$\rho_2=(3+2w+w\alpha_4)\alpha_1-3\alpha_4$.
\begin{equation}\label{eqn28}
p=\frac{Ak_5^{n+1}t^{\frac{(n+1)\alpha_1}{1+\alpha_2}}e^{(n+1)\alpha_1T_1(t)}(\rho_1+\rho_2t)^{n+1}-B(1+t)^{n+1}\left[(1-\alpha_4)t+(\alpha_2+\alpha_4)ln(1+t)\right]^{2n+2}}{k_5^{n}t^{\frac{n\alpha_1}{1+\alpha_2}}e^{n\alpha_1T_1(t)}(\rho_1+\rho_2t)^{n}(1+t)\left[(1-\alpha_4)t+(\alpha_2+\alpha_4)ln(1+t)\right]^2}.
\end{equation}
\begin{equation}\label{eqn29}
\Pi=\frac{\splitfrac{k_5^{n+1}t^{\frac{(n+1)\alpha_1}{1+\alpha_2}}e^{(n+1)\alpha_1T_1(t)}\left[\frac{\rho_1+\rho_2t}{(1+t)\left[(1-\alpha_4)t+(\alpha_2+\alpha_4)ln(1+t)\right]^2}\right]^n\times}{\left[\frac{\Pi_1+\Pi_2t}{(1+t)\left[(1-\alpha_4)t+(\alpha_2+\alpha_4)ln(1+t)\right]^2}+\frac{2k}{c_2^2t^{\frac{2}{1+\alpha_2}}e^{2T_1(t)}}\right]+B}}{k_5^{n}t^{\frac{n\alpha_1}{1+\alpha_2}}e^{n\alpha_1T_1(t)}\left[\frac{\rho_1+\rho_2t}{(1+t)\left[(1-\alpha_4)t+(\alpha_2+\alpha_4)ln(1+t)\right]^2}\right]^n},
\end{equation}
where $\Pi_1=-(1+w)\alpha_1^2-(1+3A)\alpha_2+[(\alpha_2-2)(1+A)w-1-3A+\alpha_2]\alpha_1+2$ and
$\Pi_2=-(1+w)\alpha_1^2+(1+3A)\alpha_4+[(\alpha_2-2)(1+A)w-1-3A-\alpha_4]\alpha_1+2$.
\begin{equation}\label{eqn30}
\Lambda=\frac{\Lambda_1+\Lambda_2t}{(1+t)\left[(1-\alpha_4)t+(\alpha_2+\alpha_4)ln(1+t)\right]^2}+\frac{3k}{c_2^2t^{\frac{2}{1+\alpha_2}}e^{2T_1(t)}},
\end{equation}
where $\Lambda_1=-0.5w\alpha_1^2+w(\alpha_2-2)\alpha_1-3(\alpha_2-1)$ and $\Lambda_2=-0.5w\alpha_1^2-w(\alpha_4+2)\alpha_1+3(\alpha_4+1)$.
\begin{figure}[!htb]
    \centering
    \begin{minipage}{.5\textwidth}
        \centering
        \includegraphics[width=0.8\linewidth, height=0.25\textheight]{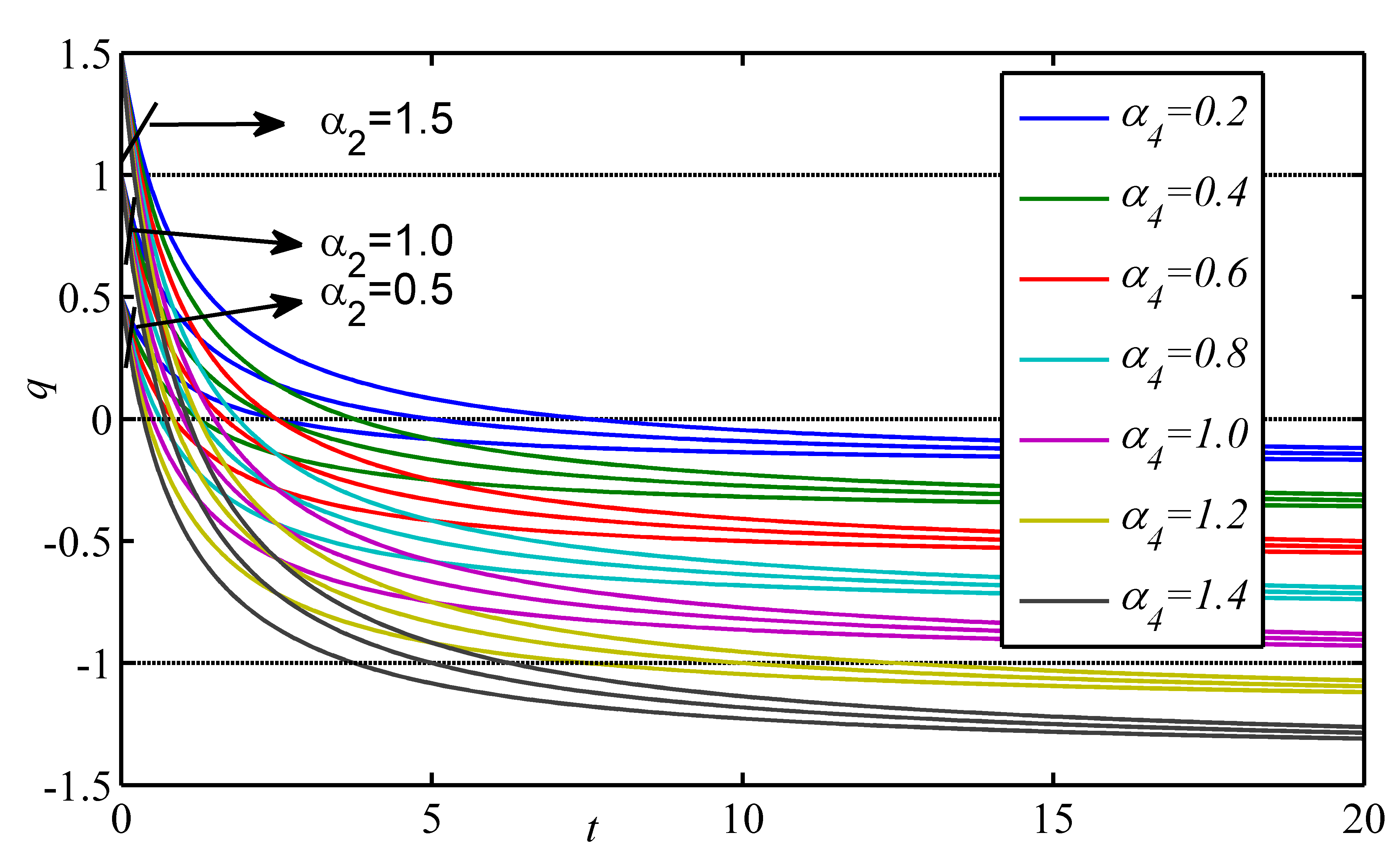}
        \caption{Variation of deceleration parameter against time for fixed $\alpha_2$ and different $\alpha_4$}
        \label{fig1}
    \end{minipage}%
    \begin{minipage}{0.5\textwidth}
        \centering
        \includegraphics[width=0.8\linewidth, height=0.25\textheight]{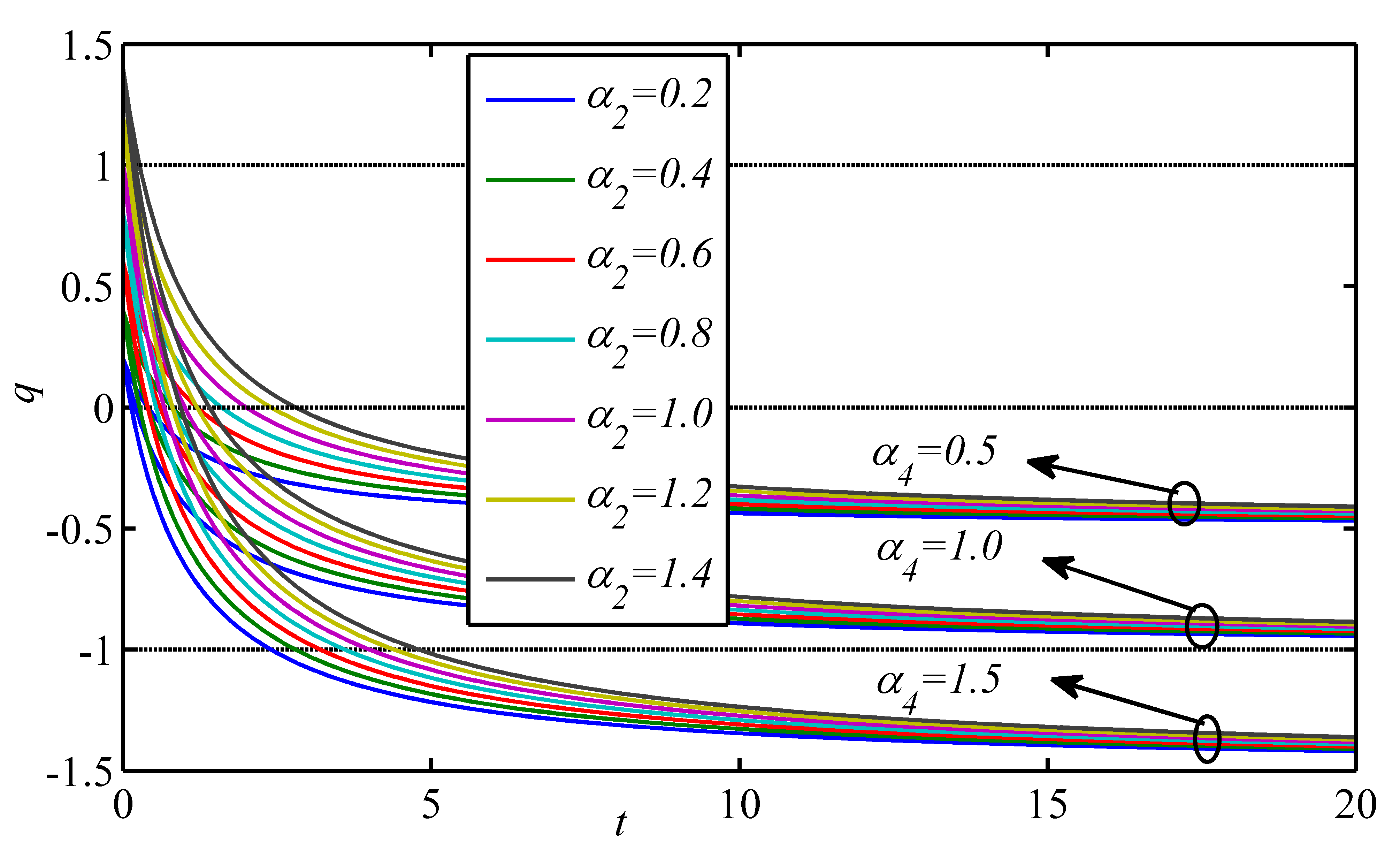}
        \caption{Variation of deceleration parameter against time for fixed $\alpha_4$ and different $\alpha_2$}
        \label{fig2}
    \end{minipage}
\end{figure}
\begin{figure}[!htb]
    \centering
    \begin{minipage}{.5\textwidth}
        \centering
        \includegraphics[width=0.8\linewidth, height=0.25\textheight]{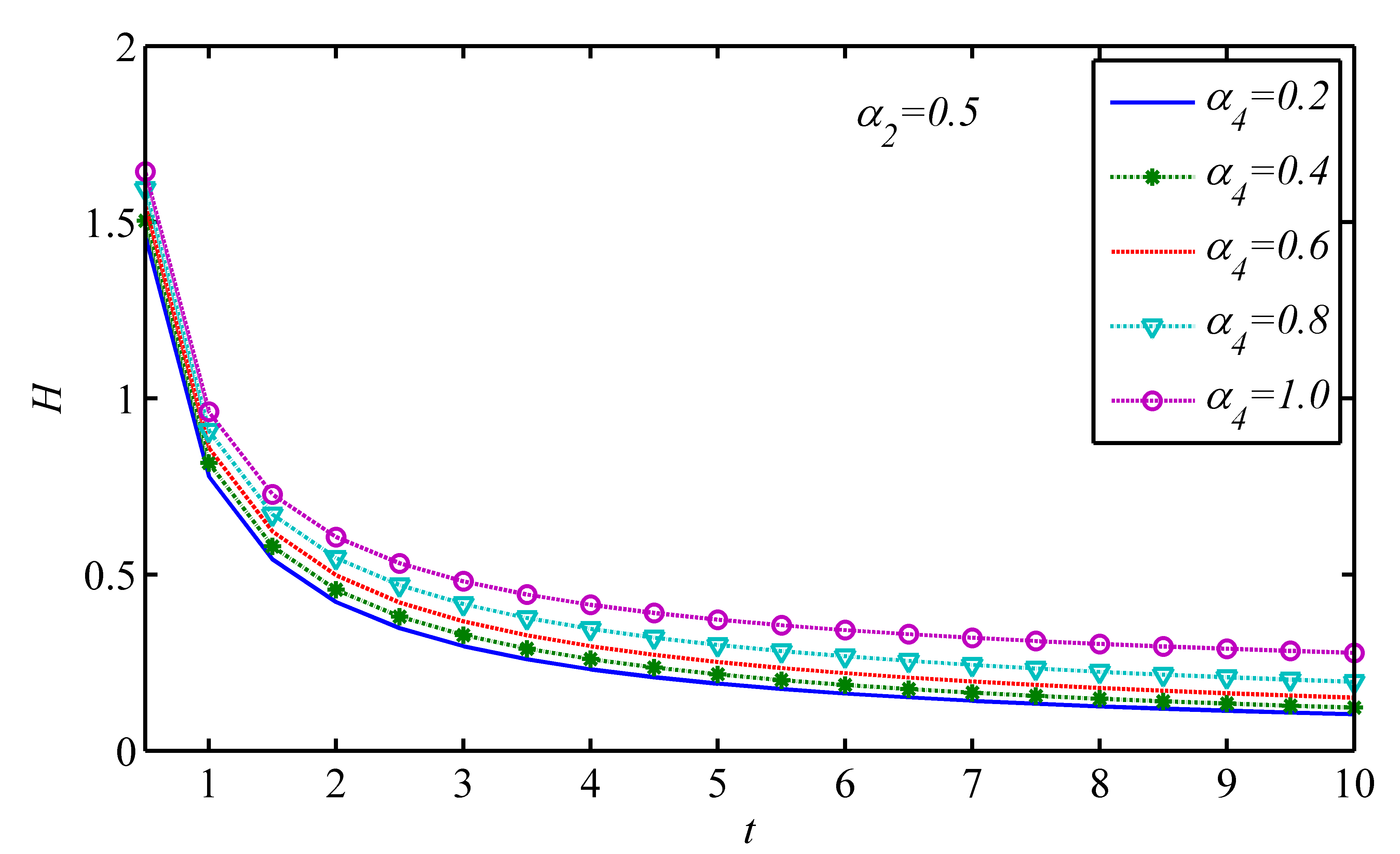}
        \caption{Variation of Hubble parameter against time  for fixed $\alpha_2=0.5$ and different $\alpha_4$}
        \label{fig3}
    \end{minipage}%
    \begin{minipage}{0.5\textwidth}
        \centering
        \includegraphics[width=0.8\linewidth, height=0.25\textheight]{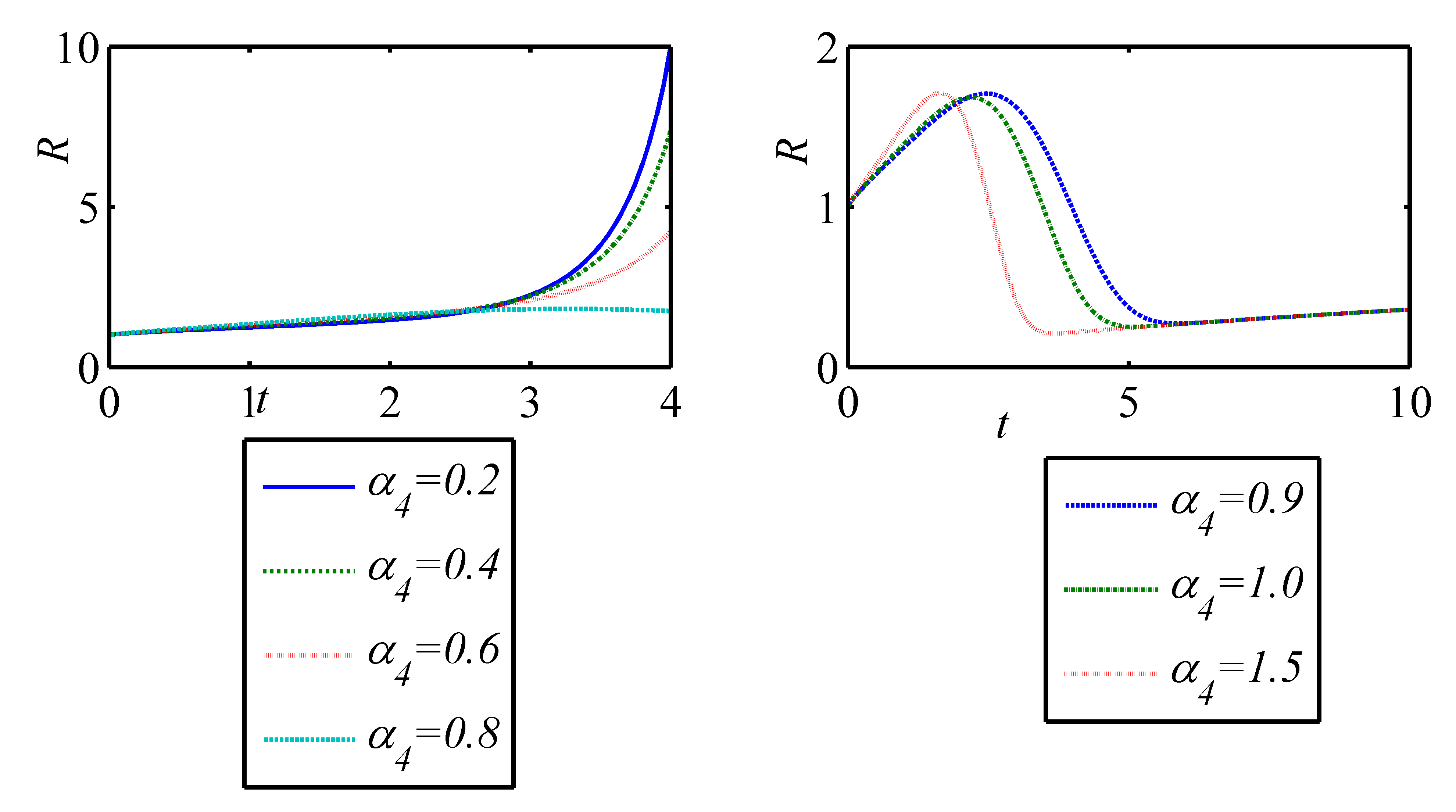}
        \caption{Variation of scale factor against time  for fixed $\alpha_2=0.5$ and different $\alpha_4$}
        \label{fig4}
    \end{minipage}
\end{figure}
Figure \ref{fig1} and Figure \ref{fig2} represents the variation of deceleration parameter against time with different values of parameters as presented in the figures for model-I. From these figures, we have noticed that when  $\alpha_2$ is fixed and  $\alpha_4$ is different and vice versa, deceleration parameter $q$ is a decreasing function of time and it takes values from positive to negative, which shows that our Universe undergoes a phase transition from deceleration phase to acceleration phase. Here we observed that $-1<q<1$ for $0<\alpha_2\leq 1$ and $0<\alpha_4\leq 1$, which means that with in the provided range of $\alpha_i (i=2,4)$ our Universe undergoes an accelerating power expansion. It can be observed from Figure \ref{fig1} and Figure \ref{fig2}.
\par
The variation of Hubble parameter $H$ and scale factor $R$ against time is plotted in the Figure \ref{fig3} and Figure \ref{fig4} respectively for model-I. As a representative case here we have presented the variation of $H$ and $R$ for fixed $\alpha_2=0.5$ and different $\alpha_4$ as in figures. It is found that Hubble parameter $H$ is a decreasing function of time and approaching towards zero with the evolution of time. For $0<\alpha_4\leq 0.8$ \& $\alpha_5=0.5$, the scale factor $R$ is an increasing function of time and higher the value of  $\alpha_4$ lower the value of  scale factor $R$. For $\alpha_4\geq 0.9$ \& $\alpha_2=0.5$, the scale factor takes a bounce and increases with the evolution of time (see Figure \ref{fig4}).
\clearpage
\begin{figure}[!htb]
  \centering
  \begin{minipage}{.5\textwidth}
  \includegraphics[width=0.8\linewidth, height=0.25\textheight]{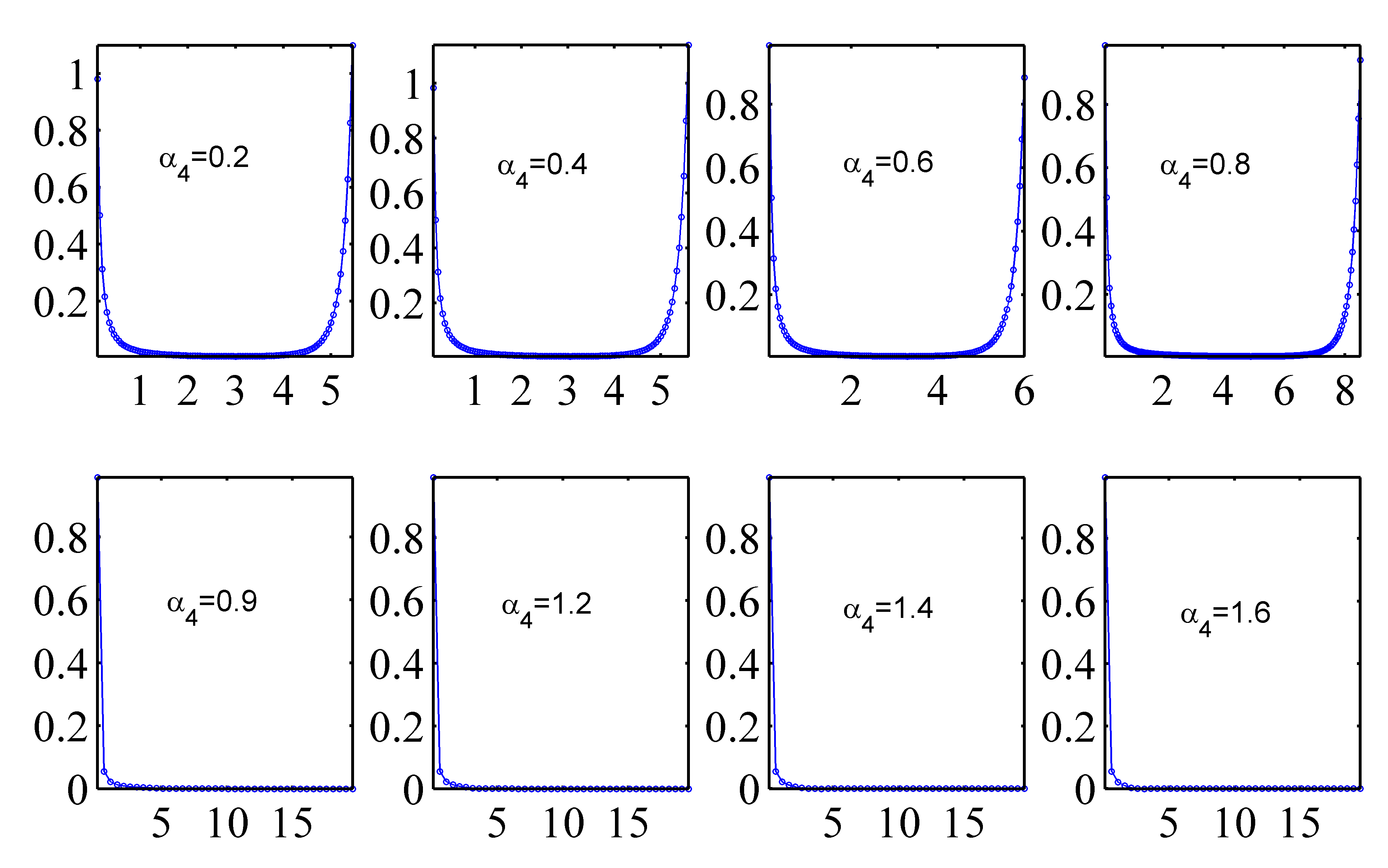}
  \caption{Variation of energy density against time  for $\alpha_1=0.5$, $\alpha_2=0.5$, $\omega=1$, $\phi_0=1$, $c_2=0.1$ and different $\alpha_4$}
  \label{fig5}

\end{minipage}%
\begin{minipage}{0.5\textwidth}
  \centering
  \includegraphics[width=0.8\linewidth, height=0.25\textheight]{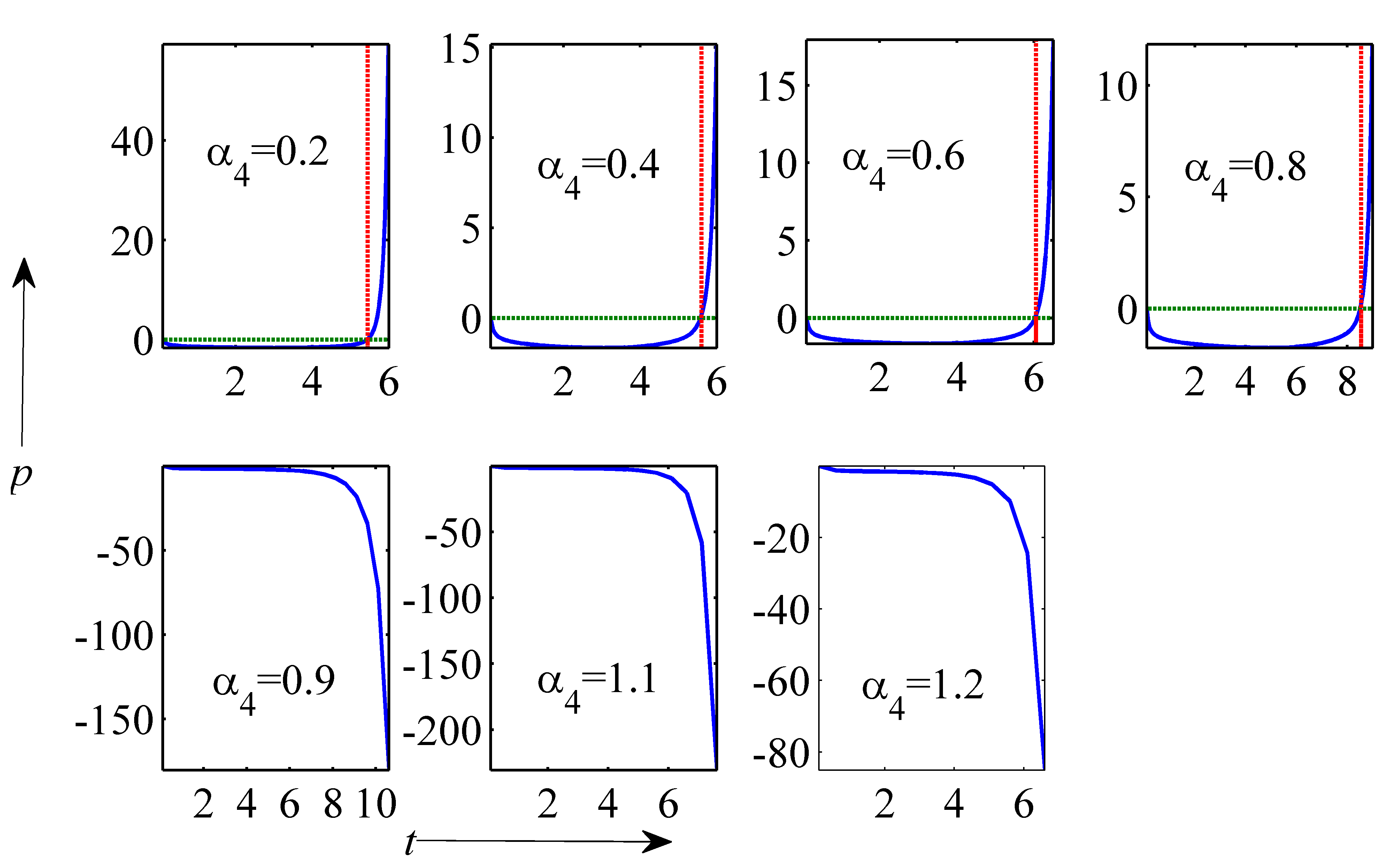}
  \caption{Variation of pressure against time  for $\alpha_1=0.5$, $\alpha_2=0.5$, $A=B=1$, $n=0.1$, $\omega=1$, $\phi_0=1$, $c_2=0.1$ and  different $\alpha_4$}
  \label{fig6}
  \end{minipage}
\end{figure}
\begin{figure}[!htb]
\begin{subfigure}
  \centering
  \includegraphics[width=.5\linewidth]{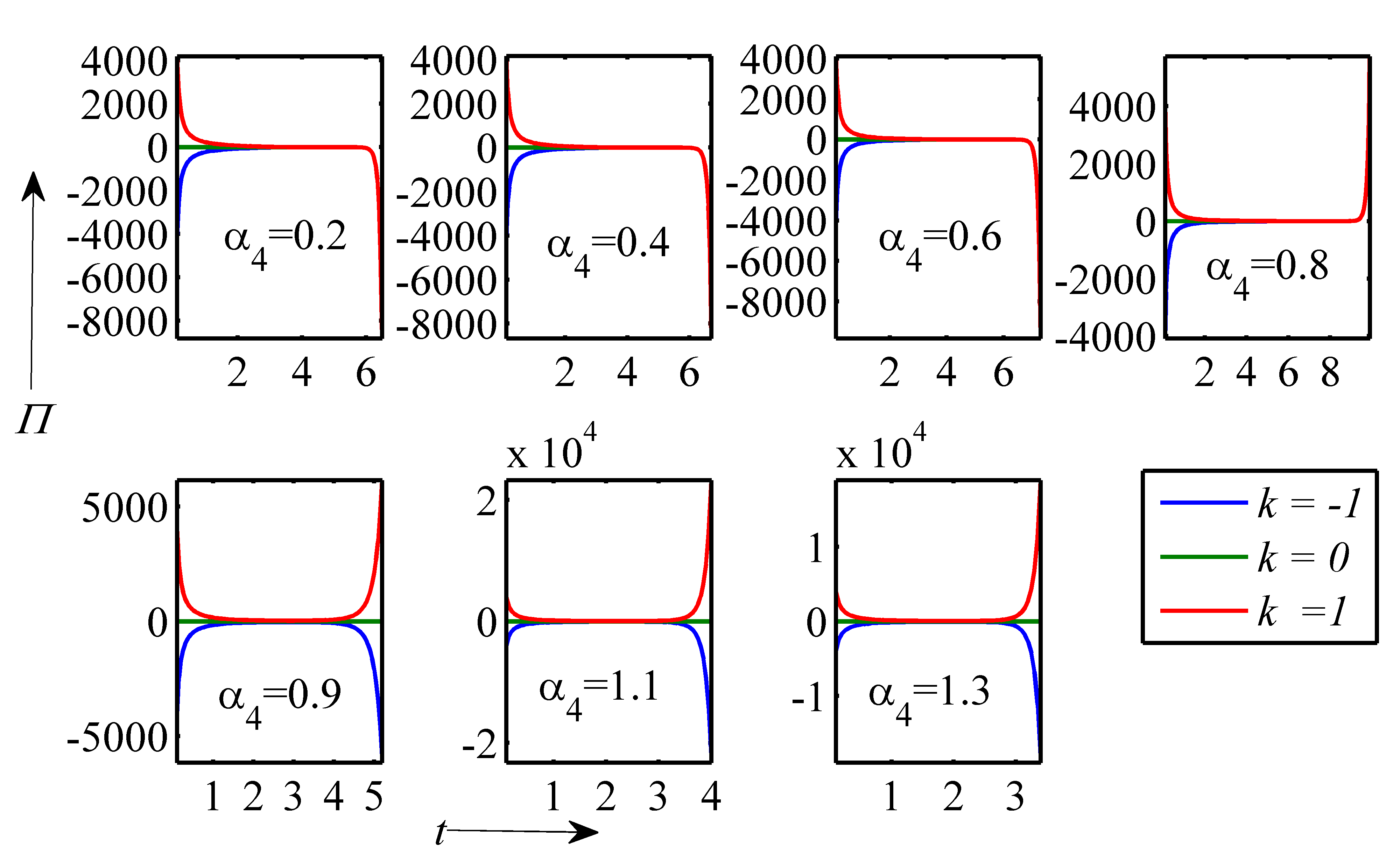}
\end{subfigure}%
\begin{subfigure}
  \centering
  \includegraphics[width=.5\linewidth]{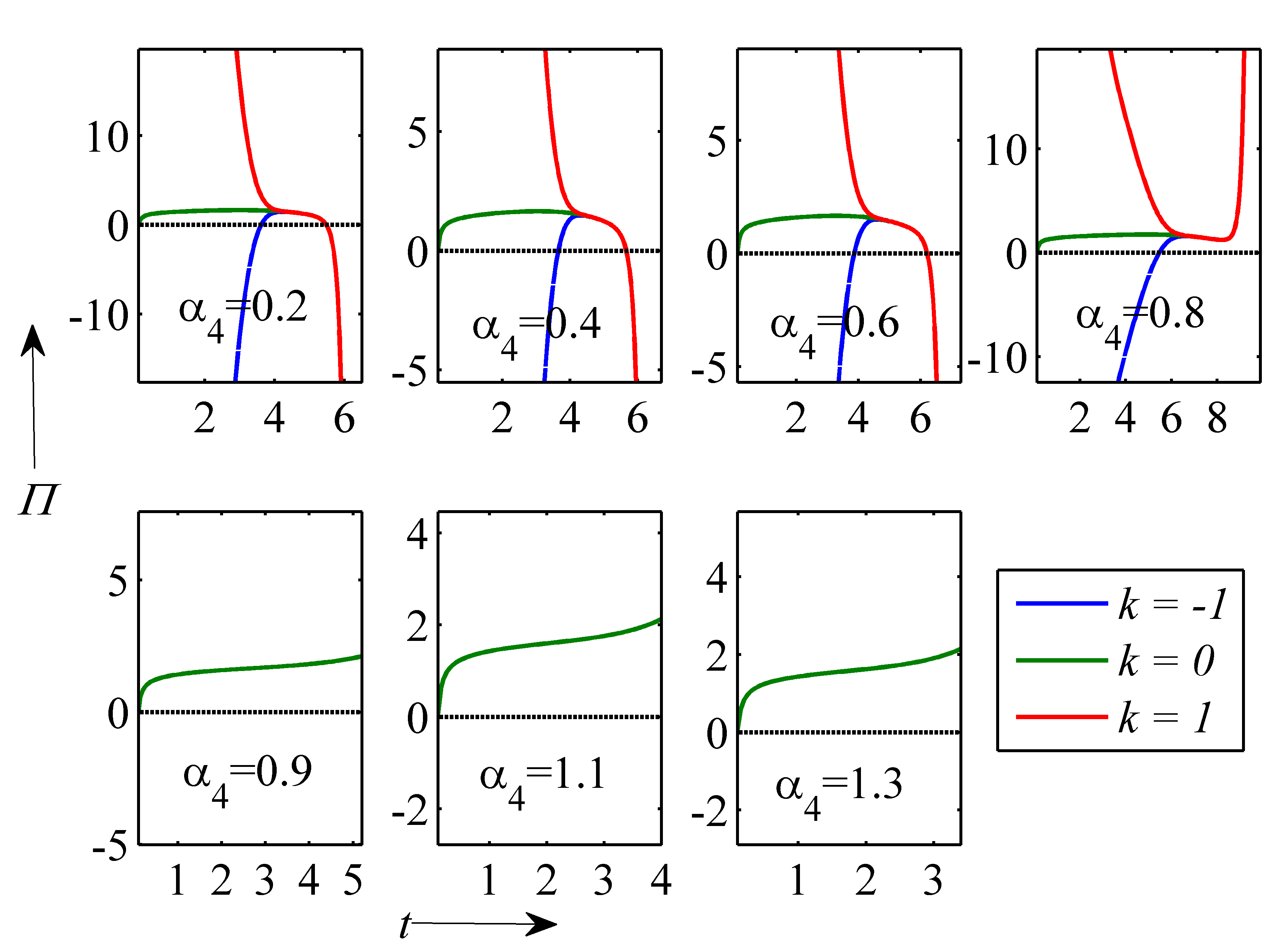}
\end{subfigure}
\caption{Variation of bulk viscous stress $\Pi$  against time  for $\alpha_1=0.5$, $\alpha_2=0.5$, $A=B=1$, $n=0.1$, $\omega=1$, $\phi_0=1$, $c_2=0.1$ and  different $\alpha_4$. Right panel shows the zooming of the left panel  figures}
\label{fig7}
\end{figure}
Figure \ref{fig5} and Figure \ref{fig6} represents the variation of energy density $\rho$ and pressure $p$ against time respectively for model-I. From the Figure \ref{fig5}  we pointed out that, in the interval $0<\alpha_4\leq 0.8$ \& $\alpha_2=0.5$ with the time, energy density decreases for small interval of time and increases to a higher value with the evolution of time. This shows that our Universe is dominated by radiation. For $\alpha_4\geq 0.9$ and $\alpha_2=0.5$ the energy density is a decreasing function of time and approaches to zero with the evolution of time. In present scenario such type of qualitative behaviour of energy density is observed from observational data. From pressure profile (Figure \ref{fig6}) we observed that, in the interval $0<\alpha_4\leq 0.8$ \& $\alpha_2=0.5$, the pressure is negative for small interval of time and increases with the evolution of time. In the interval $0.9\leq \alpha_4\leq 1.2$ \& $\alpha_2=0.5$, pressure is negative, which follow the observational data but for $\alpha_4> 1.2$ , it is complex valued, thus we neglect it.
\par
\begin{figure}[!htb]
  \centering
  \includegraphics[width=0.7\linewidth, height=0.5\textheight]{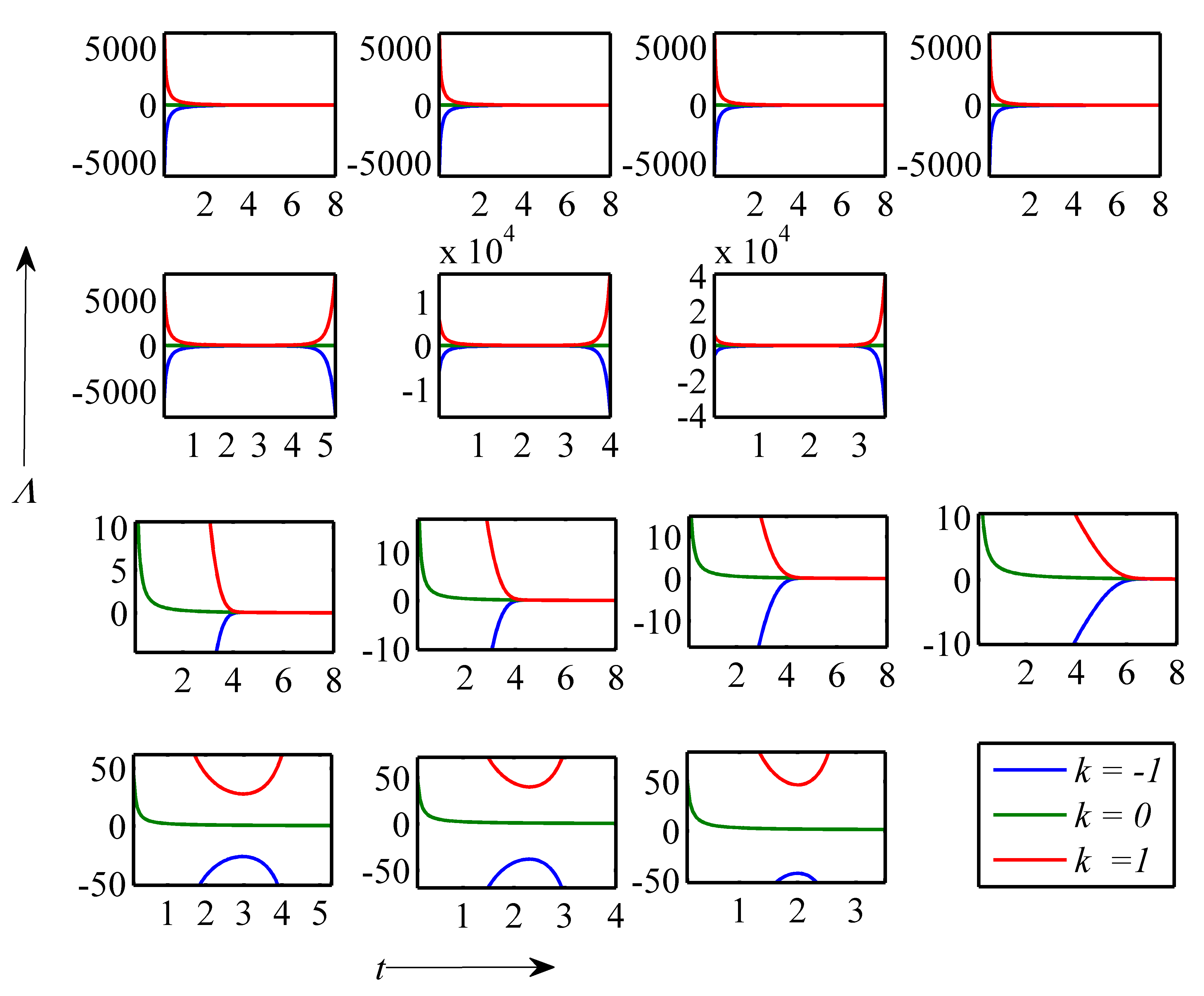}
  \caption{Variation of cosmological constant $\Lambda$ against time for  $\alpha_1=0.5$, $\alpha_2=0.5$, $\omega=1$, $c_2=0.1$ and different $\alpha_4$ $ (0.2, 0.4, 0.6, 0.8, 0.9, 1.1, 1.3)$. Lower two rows represent the magnified portion of the above two rows.}
  \label{fig8}
\end{figure}

The variation of bulk viscous stress $\Pi$ and cosmological constant $\Lambda$ against time is plotted in the Figure \ref{fig7} and Figure \ref{fig8} respectively for model-I. Figures indicate the qualitative and quantitative behaviour of both the parameters for open $(k=-1)$, flat $(k=0)$ and closed $(k=1)$ Universe. We have noticed the following points:\\ \underline{Bulk viscous stress $\Pi$ }(see Figure \ref{fig7})
\begin{itemize}
  \item Bulk viscous stress $\Pi$ takes values from positive to negative and approaches to minus infinity $(-\infty)$ with time in case of flat and closed Universe whereas negative-positive-negative valued for open Universe in the interval $0<\alpha_4\leq 0.7$ and $\alpha_2=0.5$.
  \item Bulk viscous stress $\Pi$ is positive valued and tends to infinity with the evolution of time for flat and closed Universe whereas negative-positive values for open Universe in $0.7<\alpha_4\leq 0.8$ and $\alpha_2=0.5$.
  \item For $\alpha_4>0.8$ and $\alpha_2=0.5$, bulk viscous stress $\Pi$ is positive valued and tends to infinity with the evolution of time for flat and closed Universe where as negative values for open Universe
\end{itemize}
\underline{Cosmological constant $\Lambda$}(see Figure \ref{fig8})\\
\begin{itemize}
  \item Cosmological constant $\Lambda$ is positive and negative for flat \& open Universe and closed Universe respectively. Cosmological constant $\Lambda\rightarrow 0$ when $t\rightarrow\infty$.
  \item In case of flat and  open Universe cosmological constant $\Lambda$ is positive valued for $\alpha_4>0.8$ and $\alpha_2=0.5$ whereas negative values for open Universe.
  \item In case of flat Universe cosmological constant $\Lambda\rightarrow 0$ when $t\rightarrow\infty$ but for close and open Universe $\Lambda\rightarrow \infty$ when $t\rightarrow\infty$ and $\Lambda\rightarrow -\infty$ when $t\rightarrow\infty$ respectively.
\end{itemize}
 \subsection{Model-II}
 The deceleration parameter $q$ in \eqref{eqn9} for $\alpha_2<0$ and $\alpha_3<0$ takes the form
 \begin{equation}\label{eqn31}
q=-\frac{(\alpha_5+\alpha_4 t)}{1+t},\;\;\alpha_3=-\alpha_4,\;\alpha_2=-\alpha_5,\;\textrm{and}\;\alpha_4,\alpha_5>0
 \end{equation}
Here we noticed that, $q<0$ for $\alpha_4,\alpha_5>0$, which means that our Universe is accelerating with the evolution of time.
\par
For model II, the physical parameters are obtained as follows:\\
The Hubble parameter in \eqref{eqn17} takes the form
\begin{equation}\label{eqn32}
H=\frac{1}{(1-\alpha_4)t+(\alpha_4-\alpha_5)ln(1+t)}
\end{equation}
The scale factor $R(t)$ in \eqref{eqn19} is expressed as
\begin{equation}\label{eqn33}
R=c_2t^{\frac{1}{1-\alpha_5}}e^{T_1(t)},
\end{equation}
where $T_1(t)=k_0t+k_1\frac{t^2}{2}+k_2\frac{t^3}{3}+k_3\frac{t^4}{4}+k_4\frac{t^5}{5}+O(t^6).$ and \begin{eqnarray*}
  k_0 &=& \frac{\alpha_4-\alpha_5}{2(1-\alpha_5)^2} \\
   k_1 &=&\frac{1}{1-\alpha_5}\left[\frac{(\alpha_4-\alpha_5)^2}{4(1-\alpha_5)^2}-\frac{(\alpha_4-\alpha_5)}{3(1-\alpha_5)} \right]\\
  k_2&=&\frac{1}{1-\alpha_5}\left[\frac{\alpha_4-\alpha_5}{4(1-\alpha_5)}-\frac{(\alpha_4-\alpha_5)^2}{3(1-\alpha_5)^2}+\frac{(\alpha_4-\alpha_5)^3}{8(1-\alpha_5)^3}\right] \\
  k_3&=&\frac{1}{1-\alpha_5}\left[-\frac{\alpha_4-\alpha_5}{5(1-\alpha_5)}+\frac{13(\alpha_4-\alpha_5)^2}{36(1-\alpha_5)^2}-\frac{(\alpha_4-\alpha_5)^3}{4(1-\alpha_5)^3}+\frac{(\alpha_4-\alpha_5)^4}{16(1-\alpha_5)^4}\right] \\
  k_4&=&\frac{1}{1-\alpha_5}\left[\frac{\alpha_4-\alpha_5}{6(1-\alpha_5)}-\frac{11(\alpha_4-\alpha_5)^2}{30(1-\alpha_5)^2}+\frac{17(\alpha_4-\alpha_5)^3}{48(1-\alpha_5)^3}-\frac{(\alpha_4-\alpha_5)^4}{6(1-\alpha_5)^4}\right]
\end{eqnarray*}
The FRW space-time metric in \eqref{eqn4} takes the form
\begin{equation*}
ds^2=dt^2-c_2^2t^{\frac{2}{1-\alpha_5}}e^{2T_1(t)}\left[\frac{dr^2}{1-kr^2}+r^2\left(d\theta^2+\sin^2\theta d\phi^2\right)\right]
\end{equation*}
with the above mentation $k_i$, $(i=0,1,2,3,4)$.
The energy density $(\rho)$, pressure $(p)$, bulk viscous stress $(\Pi)$ and cosmological constant $(\Lambda)$in \eqref{eqn20}, \eqref{eqn21}, \eqref{eqn22} and \eqref{eqn23} takes the form
\begin{equation}\label{eqn34}
\rho=\frac{k_5t^{\frac{\alpha_1}{1-\alpha_5}}e^{\alpha_1T_1(t)}(\rho_1+\rho_2t)}{(1+t)\left[(1-\alpha_4)t+(\alpha_4-\alpha_5)ln(1+t)\right]^2},
\end{equation}
where $k_5=\frac{\phi_0c_2^{\alpha_1}}{8\pi}$,\;$\rho_1=(3+2w+w\alpha_5)\alpha_1-3\alpha_5$,\;$\rho_2=(3+2w+w\alpha_4)\alpha_1-3\alpha_4$.
\begin{equation}\label{eqn35}
p=\frac{Ak_5^{n+1}t^{\frac{(n+1)\alpha_1}{1-\alpha_5}}e^{(n+1)\alpha_1T_1(t)}(\rho_1+\rho_2t)^{n+1}-B(1+t)^{n+1}\left[(1-\alpha_4)t+(\alpha_4-\alpha_5)ln(1+t)\right]^{2n+2}}{k_5^{n}t^{\frac{n\alpha_1}{1-\alpha_5}}e^{n\alpha_1T_1(t)}(\rho_1+\rho_2t)^{n}(1+t)\left[(1-\alpha_4)t+(\alpha_4-\alpha_5)ln(1+t)\right]^2}
\end{equation}
\begin{equation}\label{eqn36}
\Pi=\frac{\splitfrac{k_5^{n+1}t^{\frac{(n+1)\alpha_1}{1-\alpha_5}}e^{(n+1)\alpha_1T_1(t)}\left[\frac{\rho_1+\rho_2t}{(1+t)\left[(1-\alpha_4)t+(\alpha_4-\alpha_5)ln(1+t)\right]^2}\right]^n\times}{\left[\frac{\Pi_1+\Pi_2t}{(1+t)\left[(1-\alpha_4)t+(\alpha_4-\alpha_5)ln(1+t)\right]^2}+\frac{2k}{c_2^2t^{\frac{2}{1-\alpha_5}}e^{2T_1(t)}}\right]+B}}{k_5^{n}t^{\frac{n\alpha_1}{1-\alpha_5}}e^{n\alpha_1T_1(t)}\left[\frac{\rho_1+\rho_2t}{(1+t)\left[(1-\alpha_4)t+(\alpha_4-\alpha_5)ln(1+t)\right]^2}\right]^n},
\end{equation}
where $\Pi_1=-(1+w)\alpha_1^2+(1+3A)\alpha_5-[(\alpha_5+2)(1+A)w+1+3A+\alpha_5]\alpha_1+2$ and
$\Pi_2=-(1+w)\alpha_1^2+(1+3A)\alpha_4-[(\alpha_5+2)(1+A)w+1+3A+\alpha_4]\alpha_1+2$.
\begin{equation}\label{eqn37}
\Lambda=\frac{\Lambda_1+\Lambda_2t}{(1+t)\left[(1-\alpha_4)t+(\alpha_4-\alpha_5)ln(1+t)\right]^2}+\frac{3k}{c_2^2t^{\frac{2}{1-\alpha_5}}e^{2T_1(t)}},
\end{equation}
where $\Lambda_1=-0.5w\alpha_1^2-w(\alpha_5+2)\alpha_1+3(\alpha_5+1)$ and $\Lambda_2=-0.5w\alpha_1^2-w(\alpha_4+2)\alpha_1+3(\alpha_4+1)$.
\par
\begin{figure}[!htb]
    \centering
    \begin{minipage}{.5\textwidth}
        \centering
        \includegraphics[width=0.8\linewidth, height=0.25\textheight]{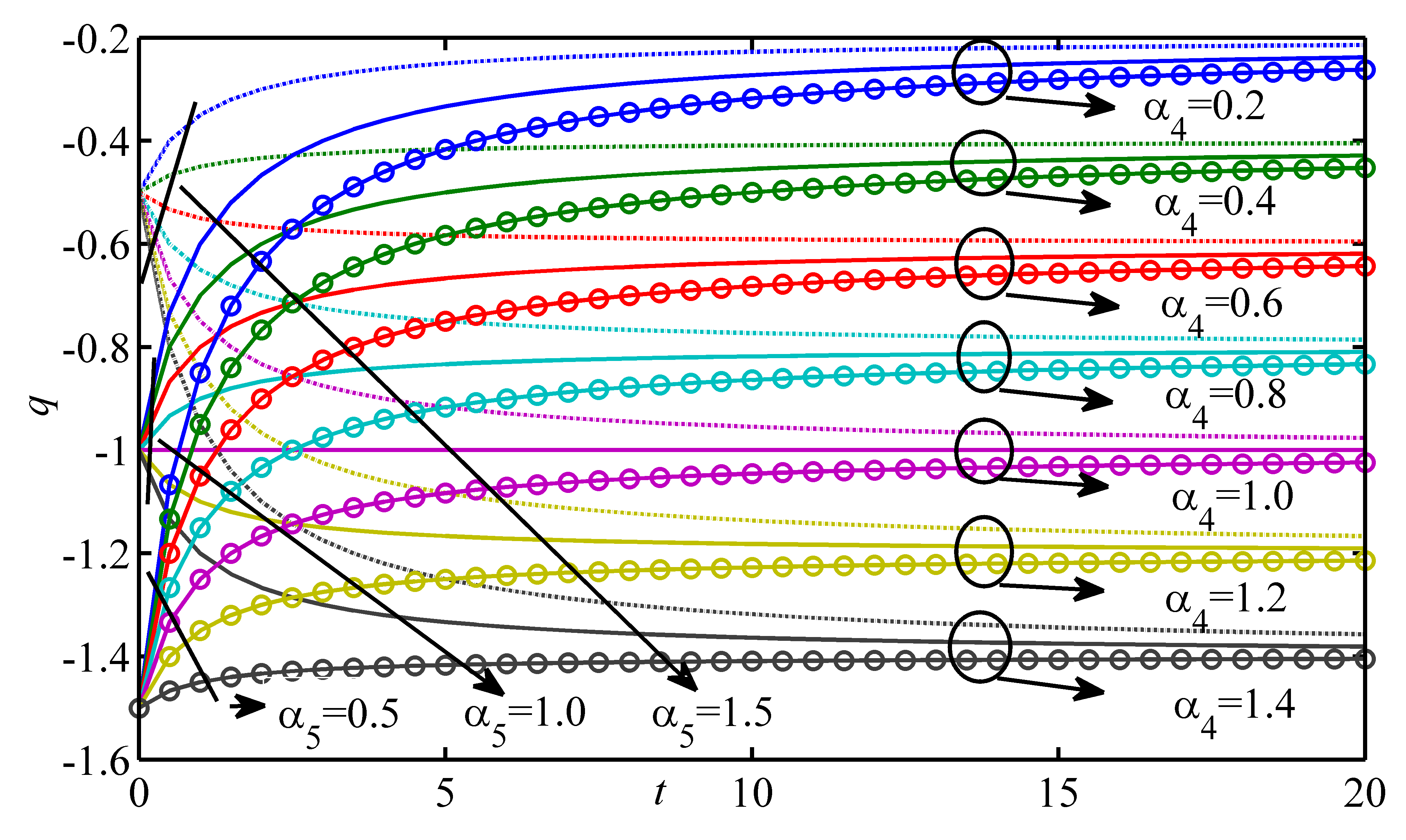}
        \caption{Variation of deceleration parameter against time for fixed $\alpha_5$ and different $\alpha_4$}
        \label{fig9}
    \end{minipage}%
    \begin{minipage}{0.5\textwidth}
        \centering
        \includegraphics[width=0.8\linewidth, height=0.25\textheight]{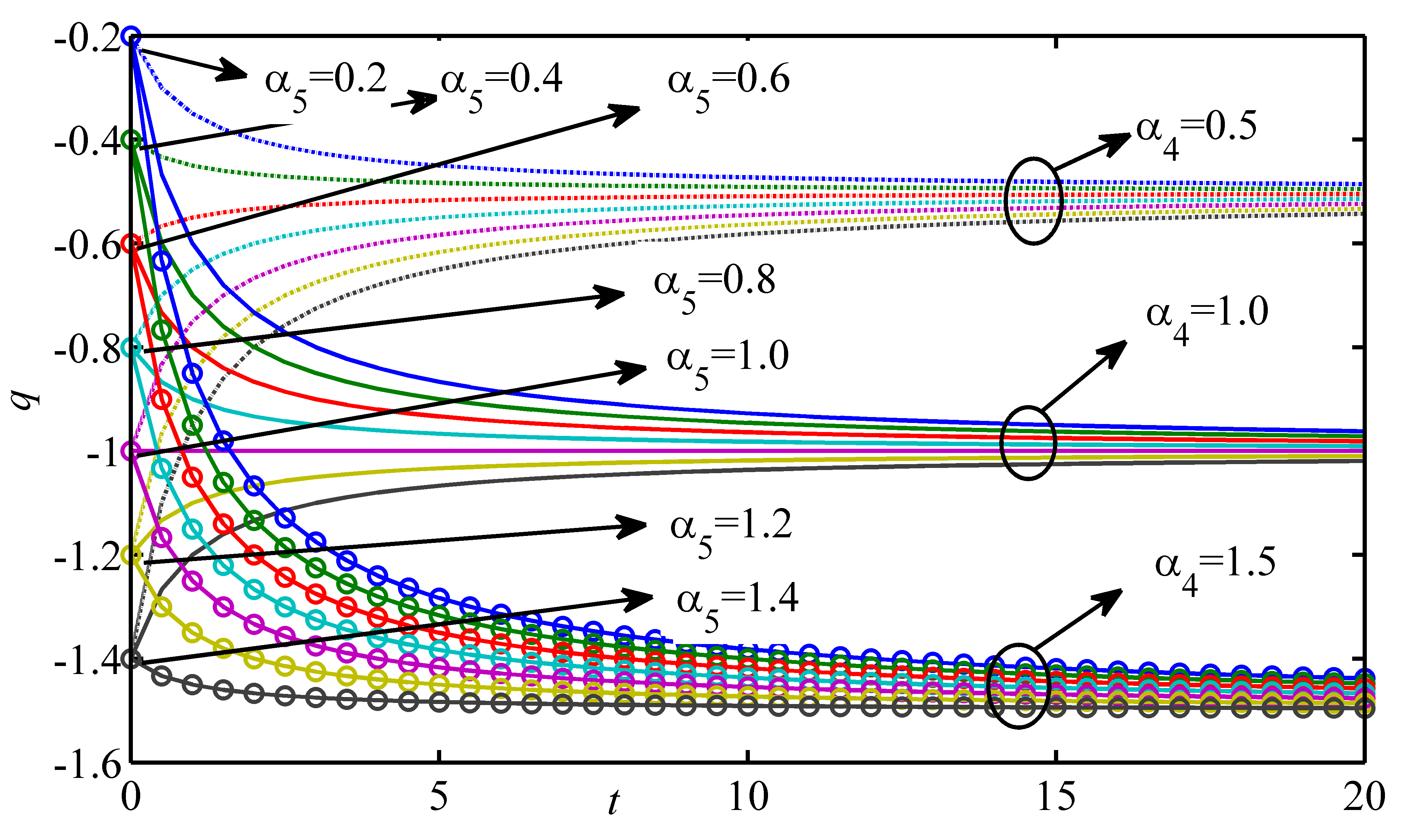}
        \caption{Variation of deceleration parameter against time  for fixed $\alpha_4$ and different $\alpha_5$}
        \label{fig10}
    \end{minipage}
\end{figure}

Now we will discuss about the physical parameters of the model-II. Figure \ref{fig9} and Figure \ref{fig10} represents the variation of deceleration parameter against time for fixed $\alpha_5$ \& different $\alpha_4$ and fixed $\alpha_4$ \& different $\alpha_5$ respectively. Here we observed that, deceleration parameter is negative and our model is accelerating.

\begin{figure}[!htb]
  \centering
  \includegraphics[width=0.5\linewidth, height=0.25\textheight]{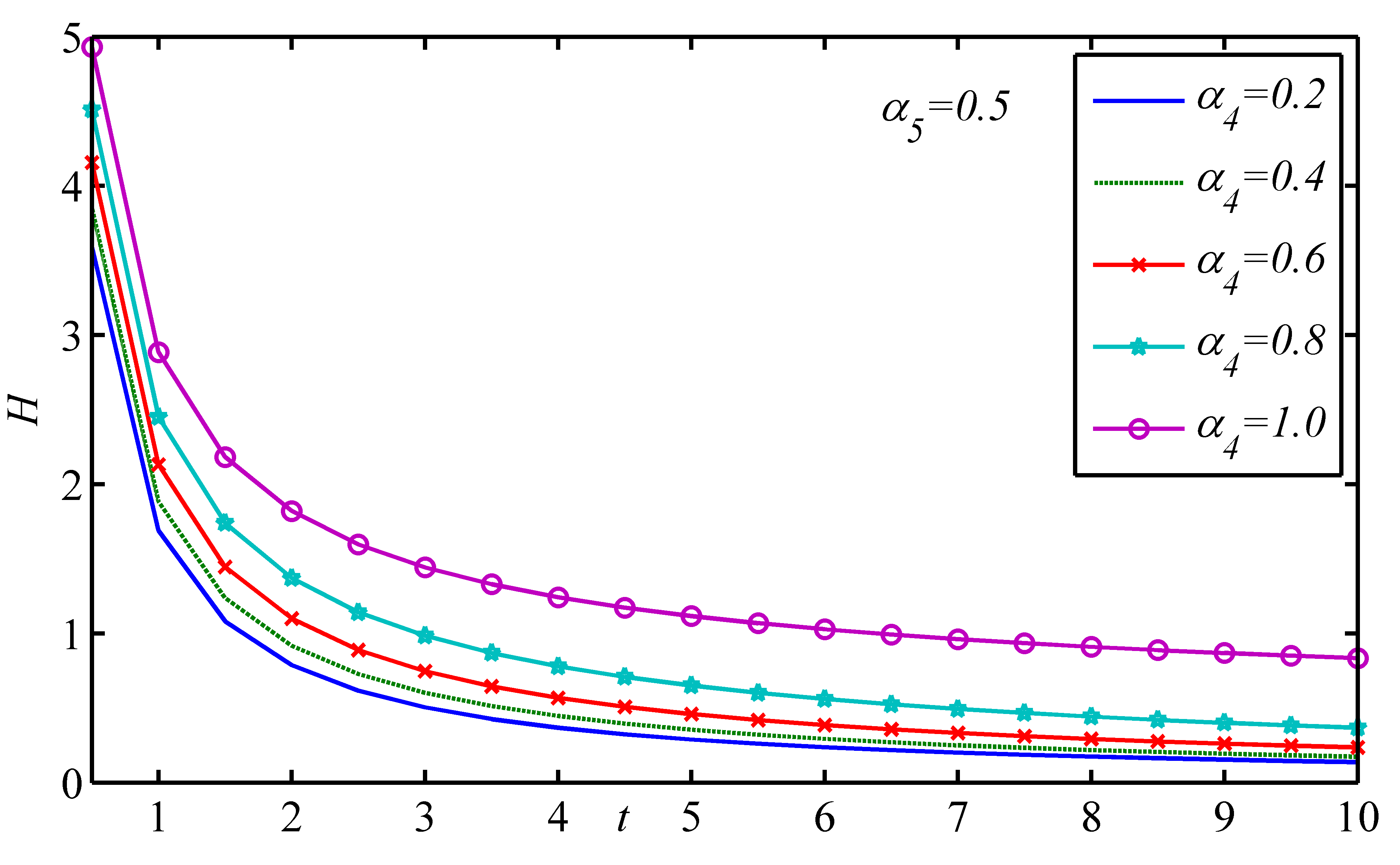}
  \caption{Variation of Hubble parameter against time  for fixed $\alpha_5=0.5$ and different $\alpha_4$}
  \label{fig11}
\end{figure}

\begin{figure}[!htb]
    \centering
    \begin{minipage}{.5\textwidth}
        \centering
        \includegraphics[width=0.8\linewidth, height=0.25\textheight]{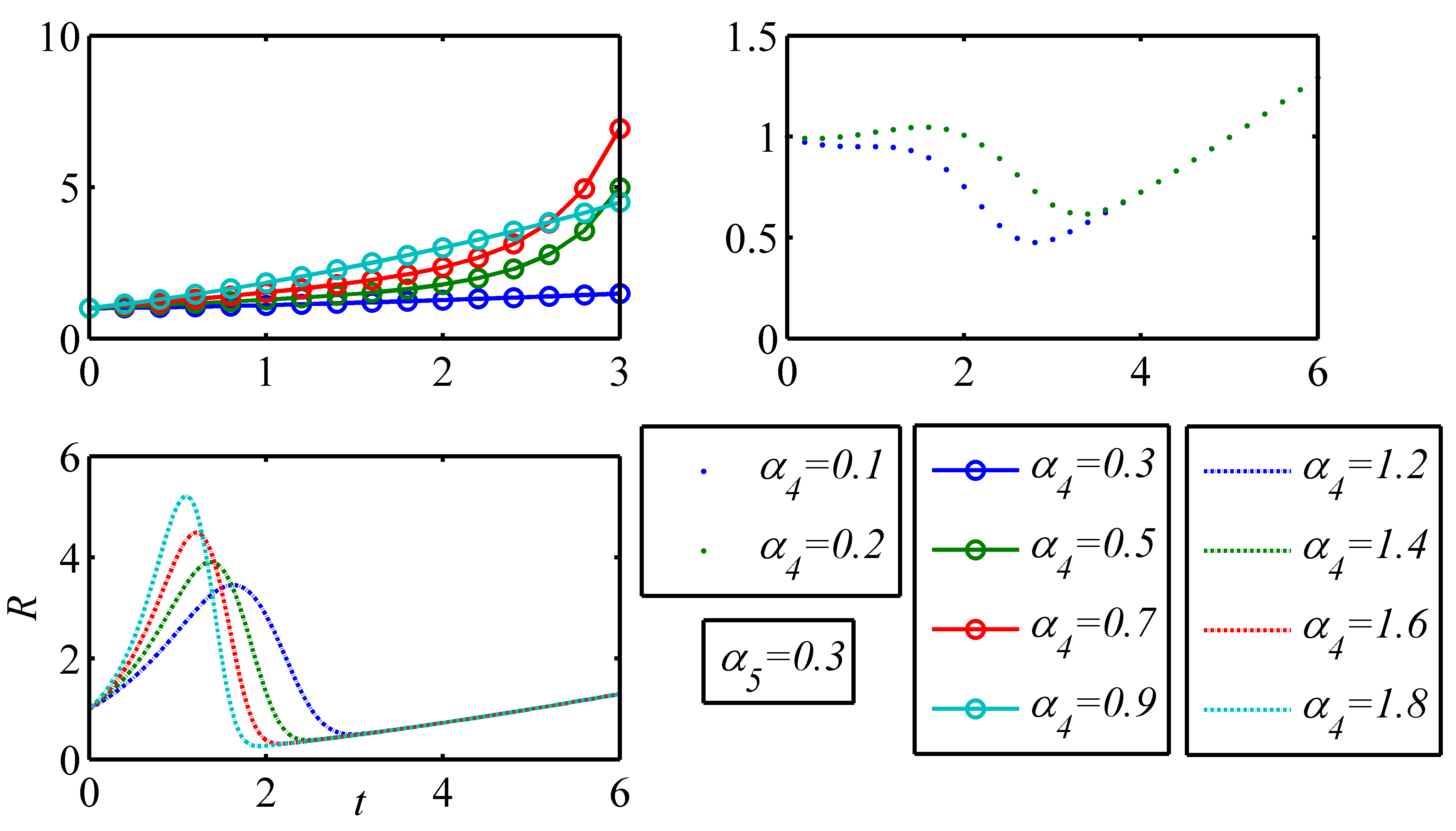}
        \caption{Variation of scale factor against time for fixed $\alpha_5=0.3$ and different $\alpha_4$}
        \label{fig12}
    \end{minipage}%
    \begin{minipage}{0.5\textwidth}
        \centering
        \includegraphics[width=0.8\linewidth, height=0.25\textheight]{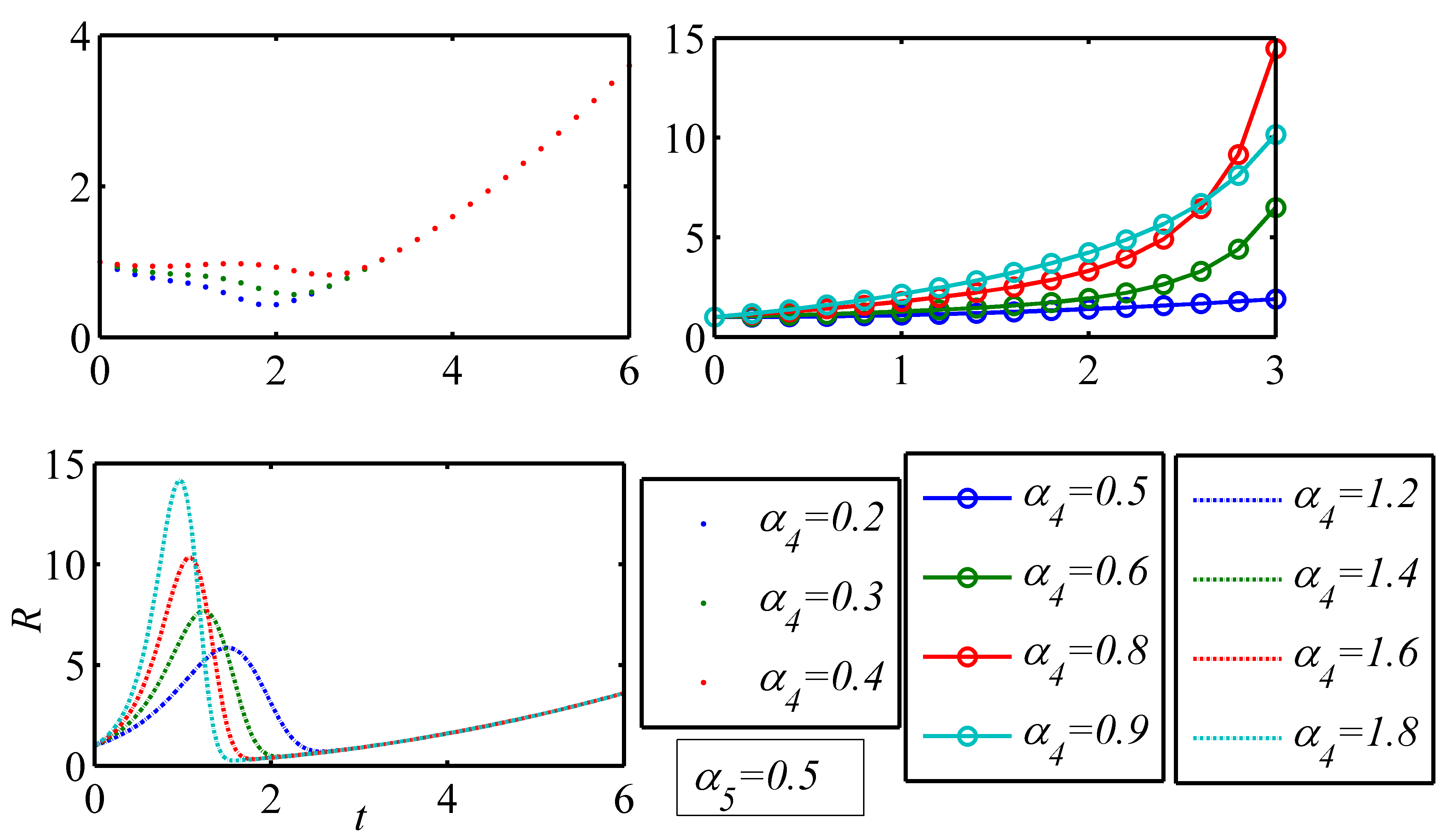}
        \caption{Variation of scale factor against time  for fixed $\alpha_5=0.5$ and different $\alpha_4$}
        \label{fig13}
    \end{minipage}
\end{figure}
Figure \ref{fig11} and Figure \ref{fig12}-\ref{fig13} depicts the variation of Hubble parameter $H$ and scale factor $R$ against time respectively for model-II. The observations are as follows:
\begin{itemize}
  \item Hubble parameter $H$ is a decreasing function of time and tending to zero with the evolution of time. As a representative case, we have presented for $\alpha_5=0.5$ and different $\alpha_4$ $(0<\alpha_4\leq 1.2)$ as in Figure \ref{fig11}.
  \item Scale factor increases with the evolution of time. Here we pointed out that, the qualitative behaviour of scale factor $R$ is different for different interval of $\alpha_4$  and $\alpha_5$. As a representative case, we choose $\alpha_5=0.3,0.5$ \& different $\alpha_4$ and all other parameters as in Figure \ref{fig12} and Figure \ref{fig13}. In the interval  $0<\alpha_4\leq 0.2$ \& $\alpha_4>0.9$ and $\alpha_5=0.3$, scale factor $R$ increases after taking a bounce where as in  $0.3\leq\alpha_4\leq 0.9$ and $\alpha_5=0.3$, it increases gradually with the evolution of time (see Figure \ref{fig12}). Similar qualitative behaviour is noticed for $\alpha_5=0.5$ and different $\alpha_4$ (see Figure \ref{fig13}).
\end{itemize}
\begin{figure}[!htb]
\begin{subfigure}
  \centering
  \includegraphics[width=.5\linewidth]{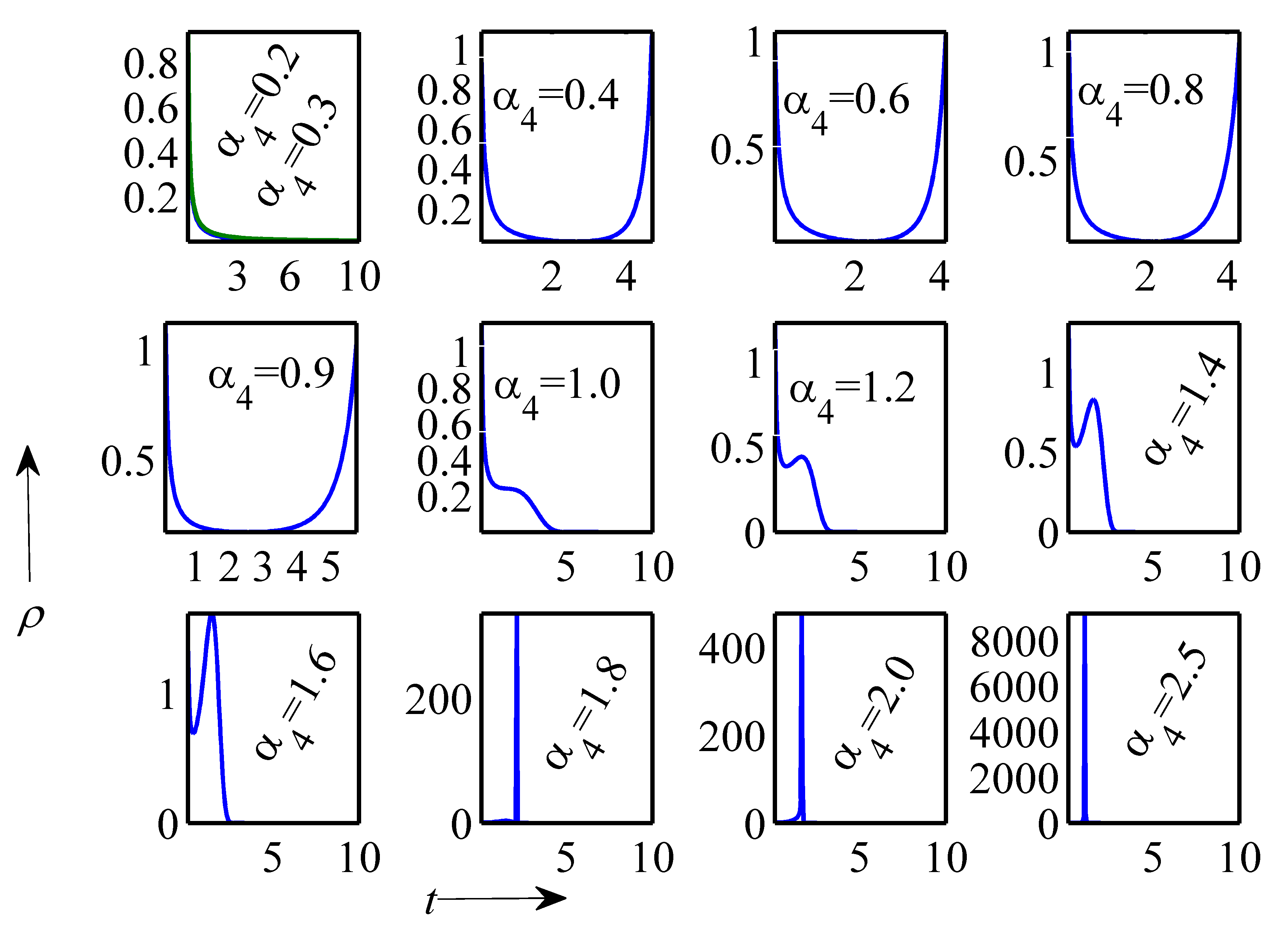}
\end{subfigure}%
\begin{subfigure}
  \centering
  \includegraphics[width=.5\linewidth]{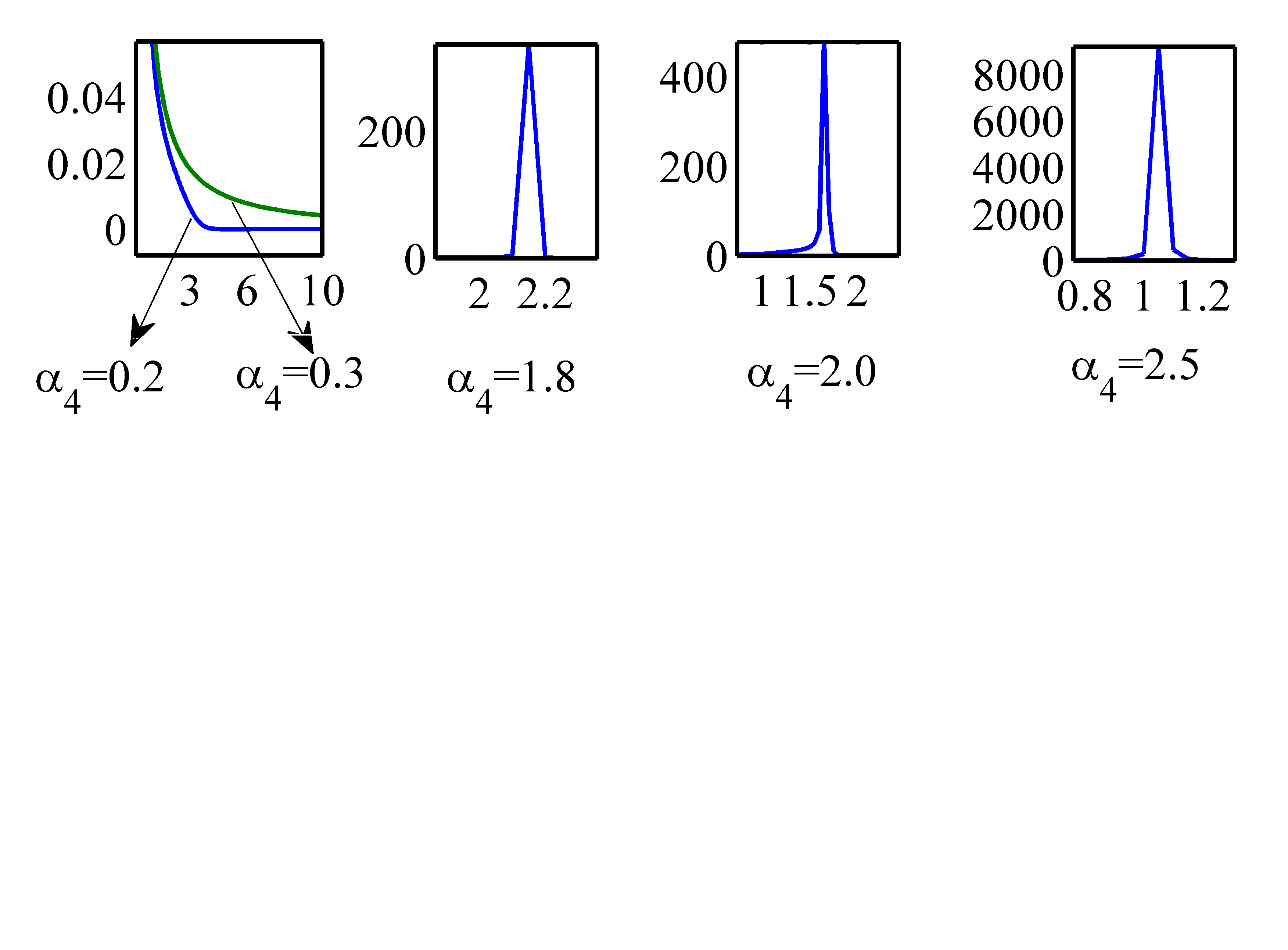}
\end{subfigure}
\caption{Variation of energy density $\rho$  against time  for $\alpha_1=0.5$, $\alpha_5=0.3$, $\omega=1$, $\phi_0=1$, $c_2=0.1$ and different $\alpha_4$. Right panel  shows the zooming of the left panel  figures.}
\label{fig14}
\end{figure}

\begin{figure}[!htb]
\begin{subfigure}
  \centering
  \includegraphics[width=.5\linewidth]{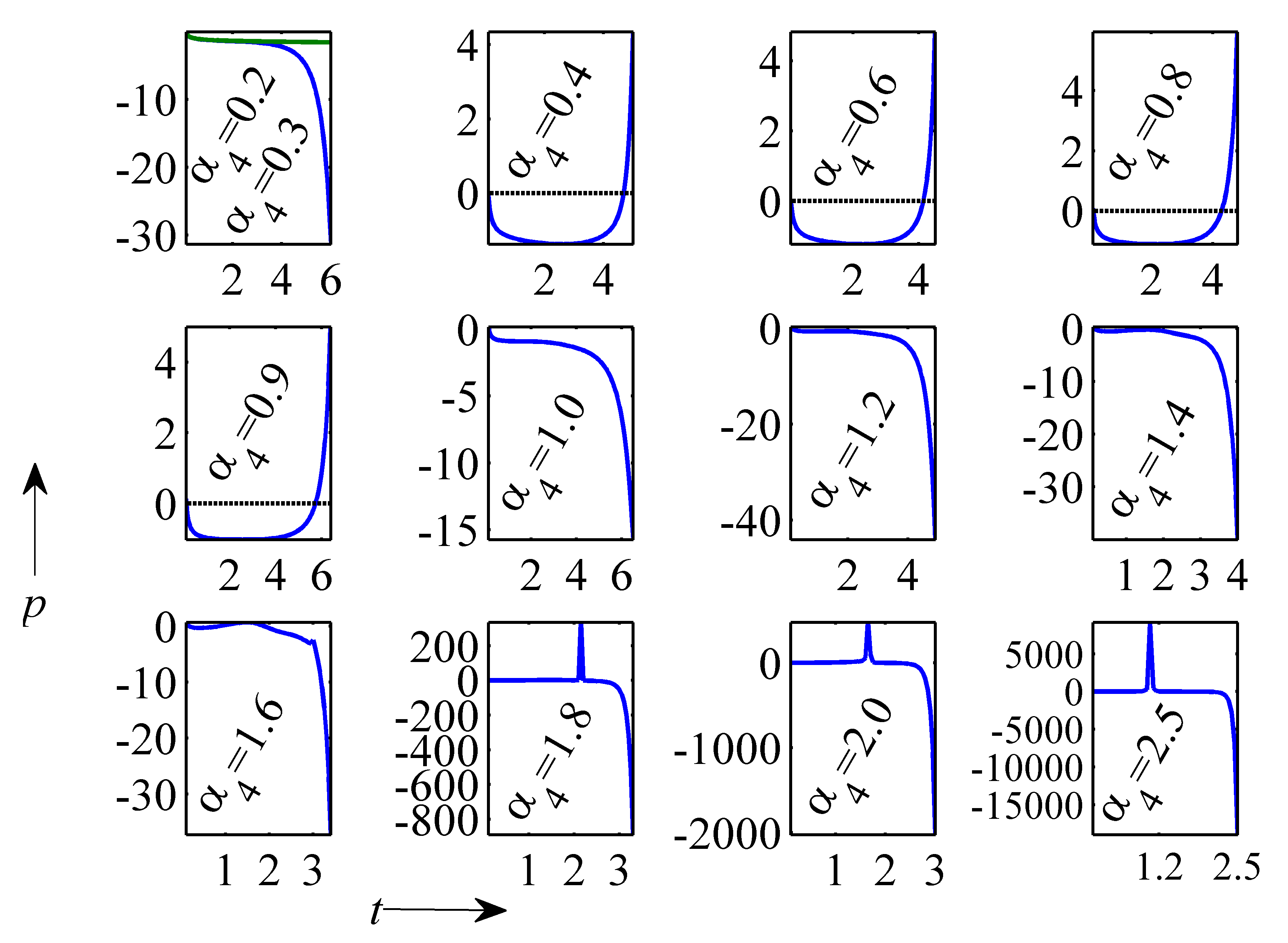}
\end{subfigure}%
\begin{subfigure}
  \centering
  \includegraphics[width=.5\linewidth]{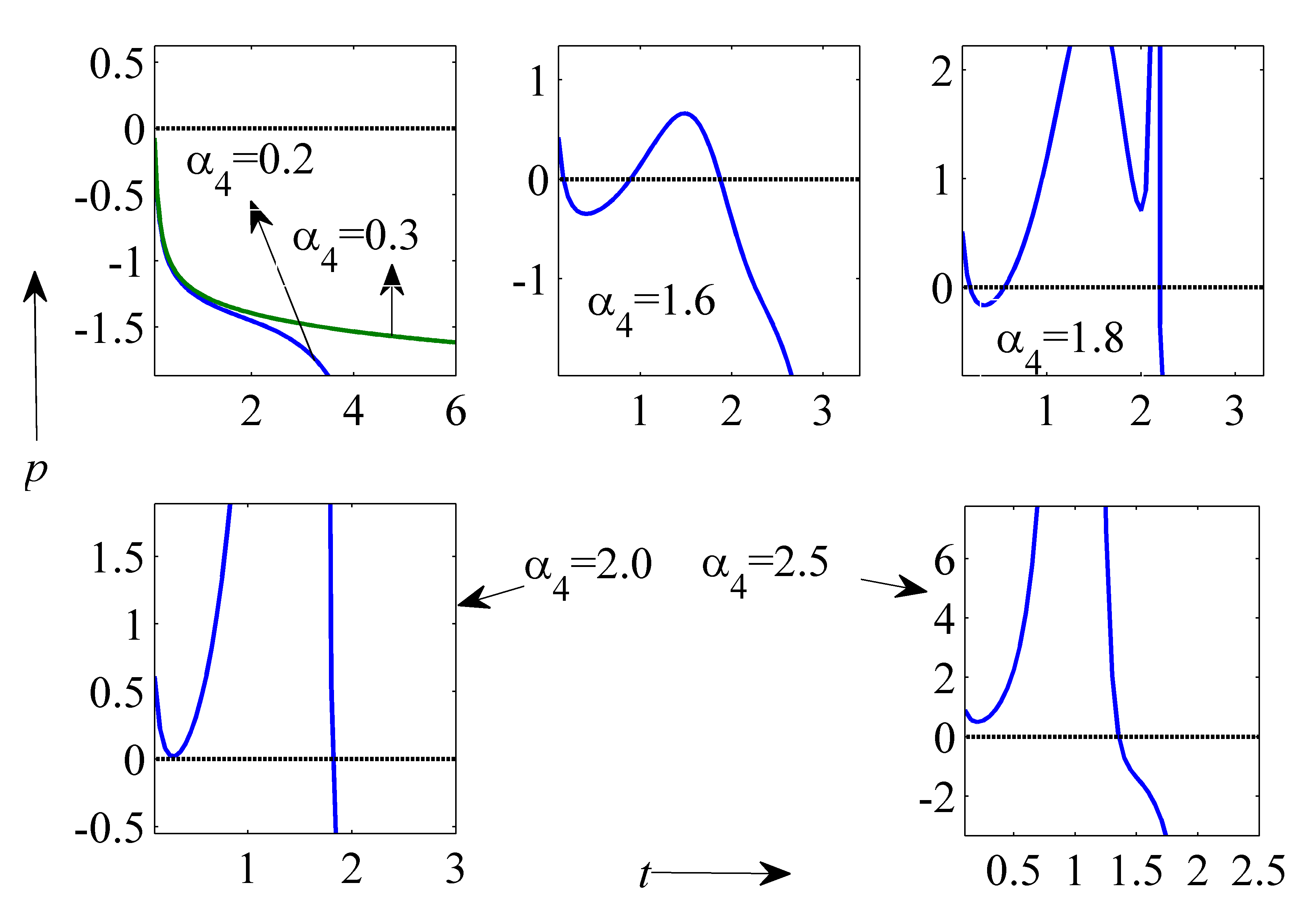}
\end{subfigure}
\caption{Variation of pressure $p$  against time  for $\alpha_1=0.5$, $\alpha_5=0.3$, $A=B=1$, $n=0.1$, $\omega=1$, $\phi_0=1$, $c_2=0.1$ different $\alpha_4$. Right panel  shows the zooming of the left panel  figures.}
\label{fig15}
\end{figure}
The variation of energy density and pressure against time is presented for model-II in the Figure \ref{fig14} and Figure \ref{fig15} respectively. As a representative case we choose $\alpha_5=0.3$ \& different $\alpha_4$ and all other parameters are as in Figure \ref{fig14} and Figure \ref{fig15}. The observations are as follows:
\begin{itemize}
  \item Energy density  gradually decreases and approaches towards zero with the evolution of time for $0<\alpha_4\leq 0.3$ and $\alpha_5=0.3$.
  \item Energy density is gradually decreased for small interval of time and tends towards infinity with the evolution of time for $0.4\leq\alpha_4\leq 0.9$ and $\alpha_5=0.3$.
  \item For $\alpha_4\geq 1$ and $\alpha_5=0.3$, energy density tends towards zero with time. Here we pointed out that, with the increment of $\alpha_4$ the bounce of the energy density increases and gradually tending to zero (see Figure \ref{fig14}).
  \item Pressure is negative in   $0<\alpha_4\leq 0.3$ and  $1\leq\alpha_4\leq 1.5$ with $\alpha_5=0.3$.
  \item Pressure is negative for a small interval of time \& gradually increases with time  and it takes values from positive to negative in the interval $0.4\leq\alpha_4\leq 0.9$ and $\alpha_4\geq 1.6$ with $\alpha_5=0.3$ respectively (see Figure \ref{fig15}).
  \end{itemize}

\begin{figure}[!htb]
\begin{subfigure}
  \centering
  \includegraphics[width=.5\linewidth]{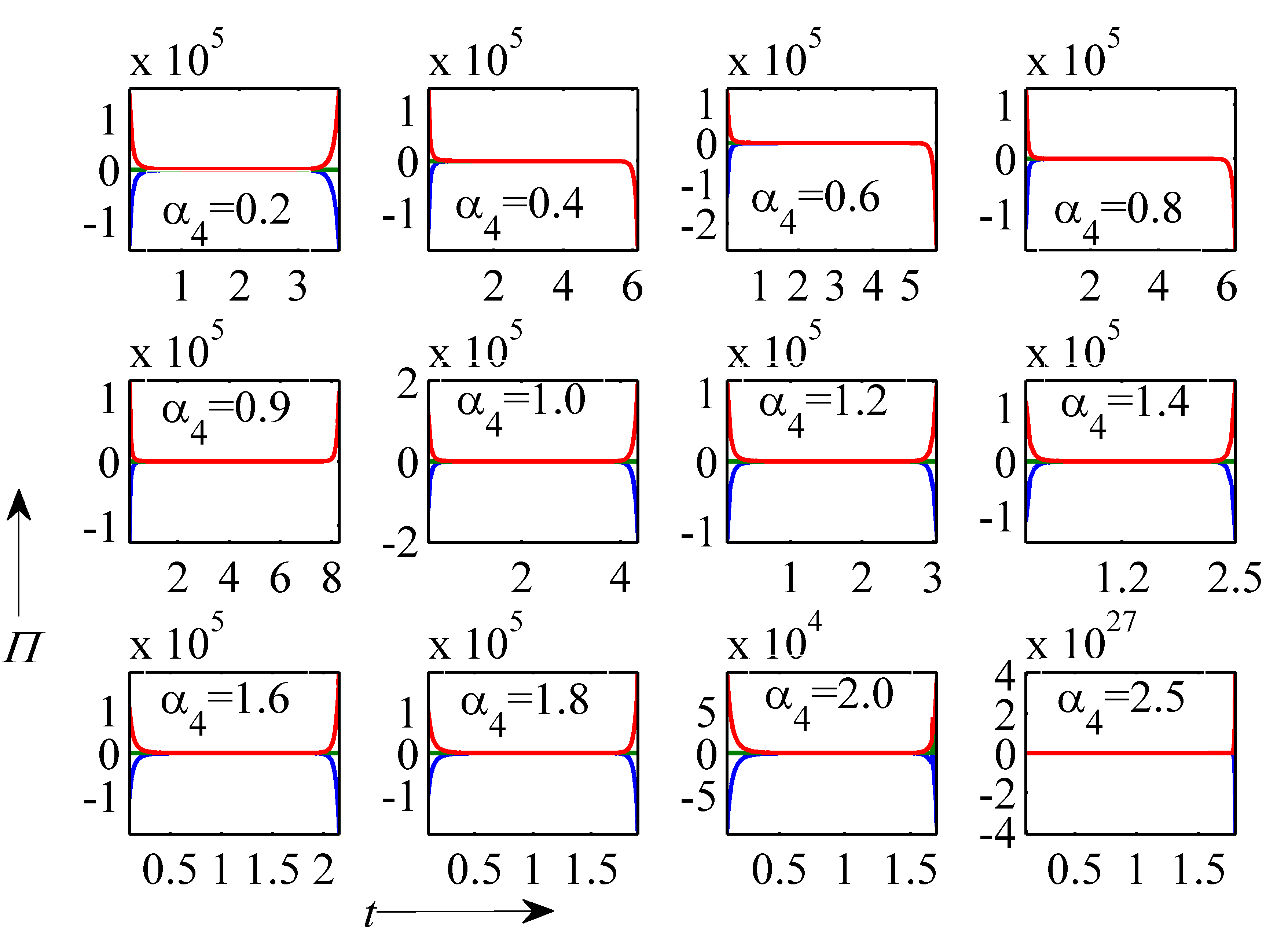}
\end{subfigure}%
\begin{subfigure}
  \centering
  \includegraphics[width=.5\linewidth]{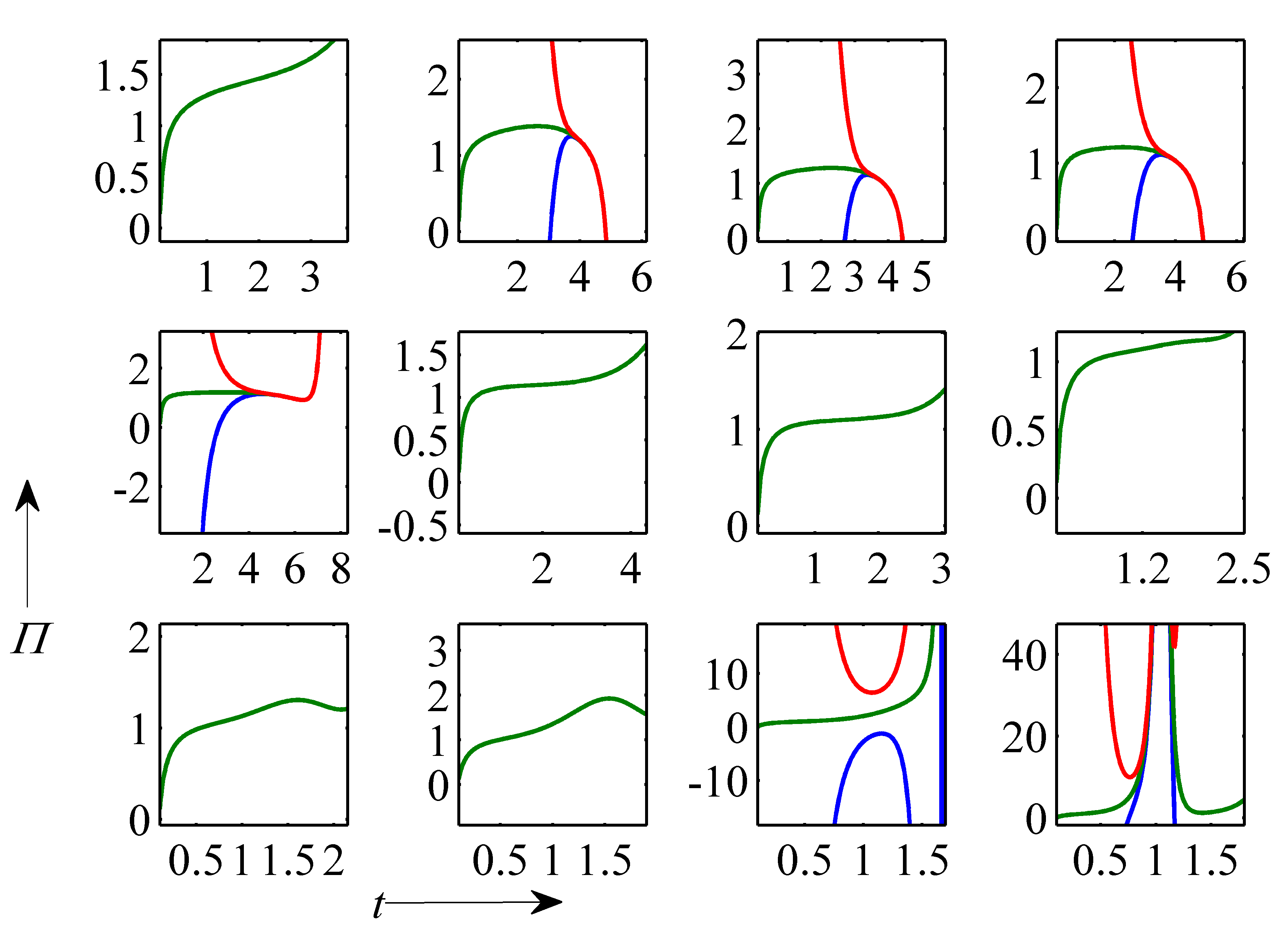}
\end{subfigure}
\caption{Variation of bulk viscous stress $\Pi$  against time  for $\alpha_1=0.5$, $\alpha_5=0.3$, $A=B=1$, $n=0.1$, $\omega=1$, $\phi_0=1$, $c_2=0.1$ and different $\alpha_4$. Right penal shows the zooming of the left penal figures. Blue line, Green line and Red line represents the $k=-1$, $k=0$ and $k=1$ respectively.}
\label{fig16}
\end{figure}

\begin{figure}[!htb]
\begin{subfigure}
  \centering
  \includegraphics[width=.5\linewidth]{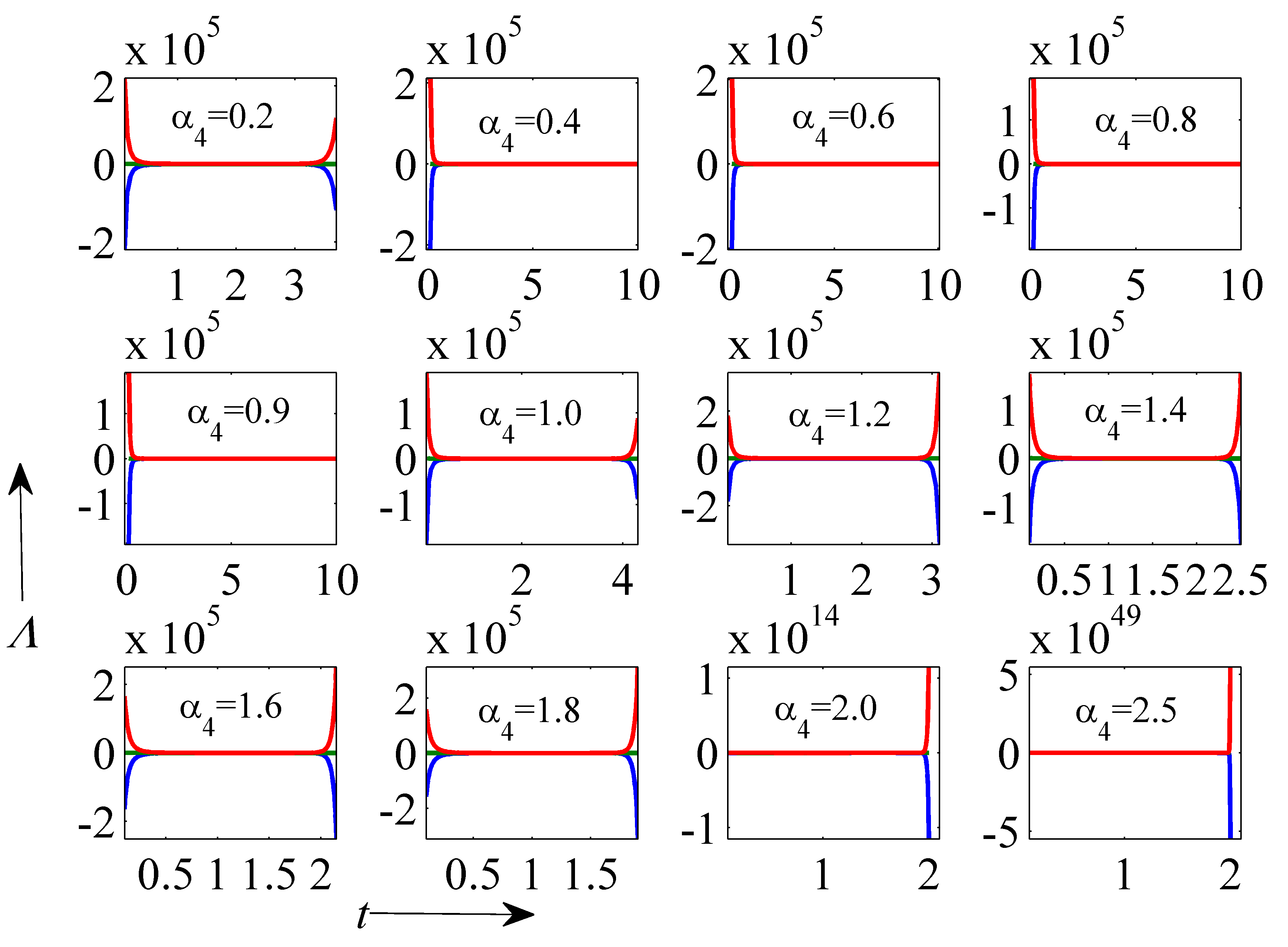}
\end{subfigure}%
\begin{subfigure}
  \centering
  \includegraphics[width=.5\linewidth]{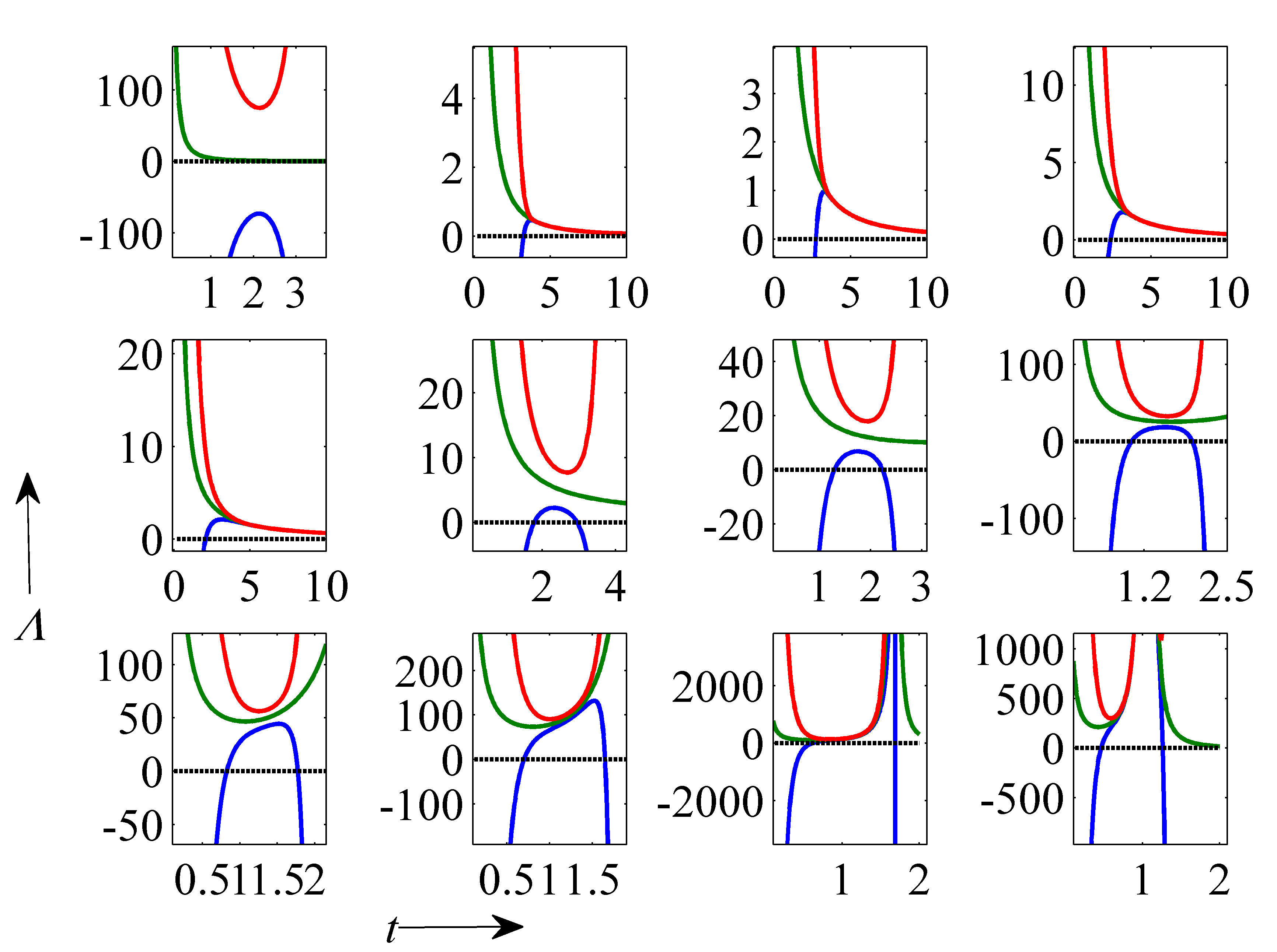}
\end{subfigure}
\caption{Variation of cosmological constant $\Lambda$  against time  for $\alpha_1=0.5$, $\alpha_5=0.3$, $\omega=1$, $c_2=0.1$  and different $\alpha_4$. Right penal shows the zooming of the left penal figures. Blue line, Green line and Red line represents the $k=-1$, $k=0$ and $k=1$ respectively.}
\label{fig17}
\end{figure}
The variation of bulk viscous stress $\Pi$ and cosmological constant $\Lambda$ against time for model-II is presented in Figure \ref{fig16} and Figure \ref{fig17} respectively. The observations are as follows:\\
\underline{Bulk viscous stress $\Pi$}(see Figure \ref{fig16})
\begin{itemize}
  \item It is positive valued for flat and closed Universe whereas negative value for open Universe in $0<\alpha_4\leq 0.2$ \& $1\leq \alpha_4\leq 2$ and $\alpha_5=0.3$.
  \item It is positive-negative valued for flat and closed Universe whereas negative-positive-negative value for open Universe in $0.3\leq \alpha_4 \leq 0.8$ and $\alpha_5=0.3$.
  \item It is positive valued for flat and closed Universe whereas negative-positive value for open Universe in $0.8< \alpha_4 \leq 1$ and $\alpha_5=0.3$. Also it approaches towards infinity with the evolution of time in the specified interval of $\alpha_4$.
  \item $\Pi$ is positive valued for flat and closed Universe whereas negative-positive-negative value for open Universe in $\alpha_4>2$ and $\alpha_5=0.3$.
\end{itemize}
\underline{Cosmological constant $\Lambda$}(see Figure \ref{fig17})
\begin{itemize}
  \item For $0<\alpha_4<0.3$ and $\alpha_5=0.3$, $\Lambda$ positive valued for flat and closed Universe whereas negative value for open Universe. In case of flat Universe $\Lambda\rightarrow 0$ when $t\rightarrow \infty$ but for open and closed Universe $\Lambda\rightarrow \infty$ and $\Lambda\rightarrow -\infty$ with time respectively.
  \item For $0.3<\alpha_4\leq 0.9$ and $\alpha_5=0.3$, $\Lambda\rightarrow 0$ when $t\rightarrow \infty$ for open, flat and closed Universe. In case of flat and closed Universe, cosmological constant is positive valued whereas in open Universe negative-positive value.
  \item It is positive valued for flat and closed Universe but it is negative-positive-negative valued for open Universe in $\alpha_4>0.9$ and $\alpha_5=0.3$.
\end{itemize}
\subsection{Model-III}
The deceleration parameter $q$ in \eqref{eqn9} for $\alpha_2<0$ and $\alpha_3=0$ takes the form
 \begin{equation}\label{eqn38}
q=-\frac{\alpha_5}{1+t},\;\;\alpha_2=-\alpha_5,\;\textrm{and}\;\alpha_5>0
 \end{equation}
Here we noticed that, $q<0$ for $\alpha_5>0$, which means that our Universe is accelerating with the evolution of time.
\par
For model III, the physical parameters are obtained as follows:\\
The Hubble parameter in \eqref{eqn17} takes the form
\begin{equation}\label{eqn39}
H=\frac{1}{t-\alpha_5ln(1+t)}
\end{equation}
The scale factor $R(t)$ in \eqref{eqn19} is expressed as
\begin{equation}\label{eqn40}
R=c_2t^{\frac{1}{1-\alpha_5}}e^{T_1(t)},
\end{equation}
where $T_1(t)=k_0t+k_1\frac{t^2}{2}+k_2\frac{t^3}{3}+k_3\frac{t^4}{4}+k_4\frac{t^5}{5}+O(t^6).$  and
\begin{eqnarray*}
  k_0 &=& -\frac{\alpha_5}{2(1-\alpha_5)^2} \\
   k_1 &=&\frac{1}{1-\alpha_5}\left[\frac{\alpha_5^2}{4(1-\alpha_5)^2}+\frac{\alpha_5}{3(1-\alpha_5)} \right]\\
  k_2&=&-\frac{1}{1-\alpha_5}\left[\frac{\alpha_5}{4(1-\alpha_5)}+\frac{\alpha_5^2}{3(1-\alpha_5)^2}+\frac{\alpha_5^3}{8(1-\alpha_5)^3}\right] \\
  k_3&=&\frac{1}{1-\alpha_5}\left[\frac{\alpha_5}{5(1-\alpha_5)}+\frac{13\alpha_5^2}{36(1-\alpha_5)^2}+\frac{\alpha_5^3}{4(1-\alpha_5)^3}+\frac{\alpha_5^4}{16(1-\alpha_5)^4}\right] \\
  k_4&=&-\frac{1}{1-\alpha_5}\left[\frac{\alpha_5}{6(1-\alpha_5)}+\frac{11\alpha_5^2}{30(1-\alpha_5)^2}+\frac{17\alpha_5^3}{48(1-\alpha_5)^3}+\frac{\alpha_5^4}{6(1-\alpha_5)^4}\right] \end{eqnarray*}
The FRW space-time metric in \eqref{eqn4} takes the form
\begin{equation*}
ds^2=dt^2-c_2^2t^{\frac{2}{1-\alpha_5}}e^{2T_1(t)}\left[\frac{dr^2}{1-kr^2}+r^2\left(d\theta^2+\sin^2\theta d\phi^2\right)\right]
\end{equation*}
with the above mentation $k_i$, $(i=0,1,2,3,4)$.
The energy density $(\rho)$, pressure $(p)$, bulk viscous stress $(\Pi)$ and cosmological constant $(\Lambda)$in \eqref{eqn20}, \eqref{eqn21}, \eqref{eqn22} and \eqref{eqn23} takes the form
\begin{equation}\label{eqn41}
\rho=\frac{k_5t^{\frac{\alpha_1}{1-\alpha_5}}e^{\alpha_1T_1(t)}(\rho_1+\rho_2t)}{(1+t)\left[t-\alpha_5ln(1+t)\right]^2},
\end{equation}
where $k_5=\frac{\phi_0c_2^{\alpha_1}}{8\pi}$,\;$\rho_1=(3+2w+w\alpha_5)\alpha_1-3\alpha_5$,\;$\rho_2=(3+2w)\alpha_1$.
\begin{equation}\label{eqn42}
p=\frac{Ak_5^{n+1}t^{\frac{(n+1)\alpha_1}{1-\alpha_5}}e^{(n+1)\alpha_1T_1(t)}(\rho_1+\rho_2t)^{n+1}-B(1+t)^{n+1}\left[t-\alpha_5ln(1+t)\right]^{2n+2}}{k_5^{n}t^{\frac{n\alpha_1}{1-\alpha_5}}e^{n\alpha_1T_1(t)}(\rho_1+\rho_2t)^{n}(1+t)\left[t-\alpha_5ln(1+t)\right]^2}
\end{equation}
\begin{equation}\label{eqn43}
\Pi=\frac{\splitfrac{k_5^{n+1}t^{\frac{(n+1)\alpha_1}{1-\alpha_5}}e^{(n+1)\alpha_1T_1(t)}\left[\frac{\rho_1+\rho_2t}{(1+t)\left[t-\alpha_5ln(1+t)\right]^2}\right]^n\times}{\left[\frac{\Pi_1+\Pi_2t}{(1+t)\left[t-\alpha_5 ln(1+t)\right]^2}+\frac{2k}{c_2^2t^{\frac{2}{1-\alpha_5}}e^{2T_1(t)}}\right]+B}}{k_5^{n}t^{\frac{n\alpha_1}{1-\alpha_5}}e^{n\alpha_1T_1(t)}\left[\frac{\rho_1+\rho_2t}{(1+t)\left[t-\alpha_5ln(1+t)\right]^2}\right]^n},
\end{equation}
where $\Pi_1=-(1+w)\alpha_1^2+(1+3A)\alpha_5-[(\alpha_5+2)(1+A)w+1+3A+\alpha_5]\alpha_1+2$ and
$\Pi_2=-(1+w)\alpha_1^2-[(\alpha_5+2)(1+A)w+1+3A]\alpha_1+2$.
\begin{equation}\label{eqn44}
\Lambda=\frac{\Lambda_1+\Lambda_2t}{(1+t)\left[t-\alpha_5ln(1+t)\right]^2}+\frac{3k}{c_2^2t^{\frac{2}{1-\alpha_5}}e^{2T_1(t)}},
\end{equation}
where $\Lambda_1=-[0.5w\alpha_1^2+w(\alpha_5+2)\alpha_1-3(\alpha_5+1)]$ and $\Lambda_2=-0.5w\alpha_1^2-2w\alpha_1+3$.

\begin{figure}[!htb]
    \centering
    \begin{minipage}{.5\textwidth}
        \centering
        \includegraphics[width=0.8\linewidth, height=0.25\textheight]{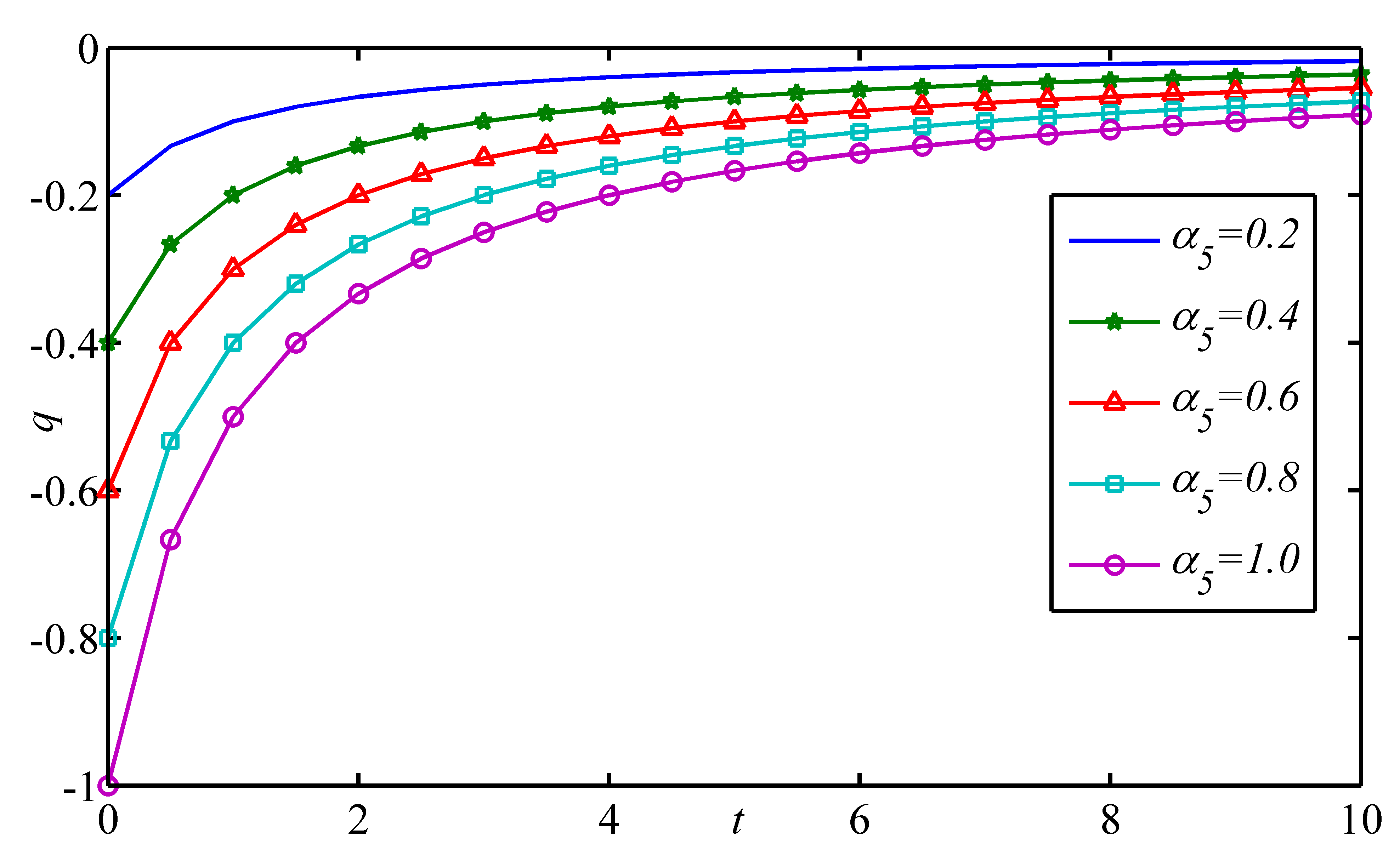}
        \caption{Variation of deceleration parameter against time\newline for different $\alpha_5$}
        \label{fig18}
    \end{minipage}%
    \begin{minipage}{0.5\textwidth}
        \centering
        \includegraphics[width=0.8\linewidth, height=0.25\textheight]{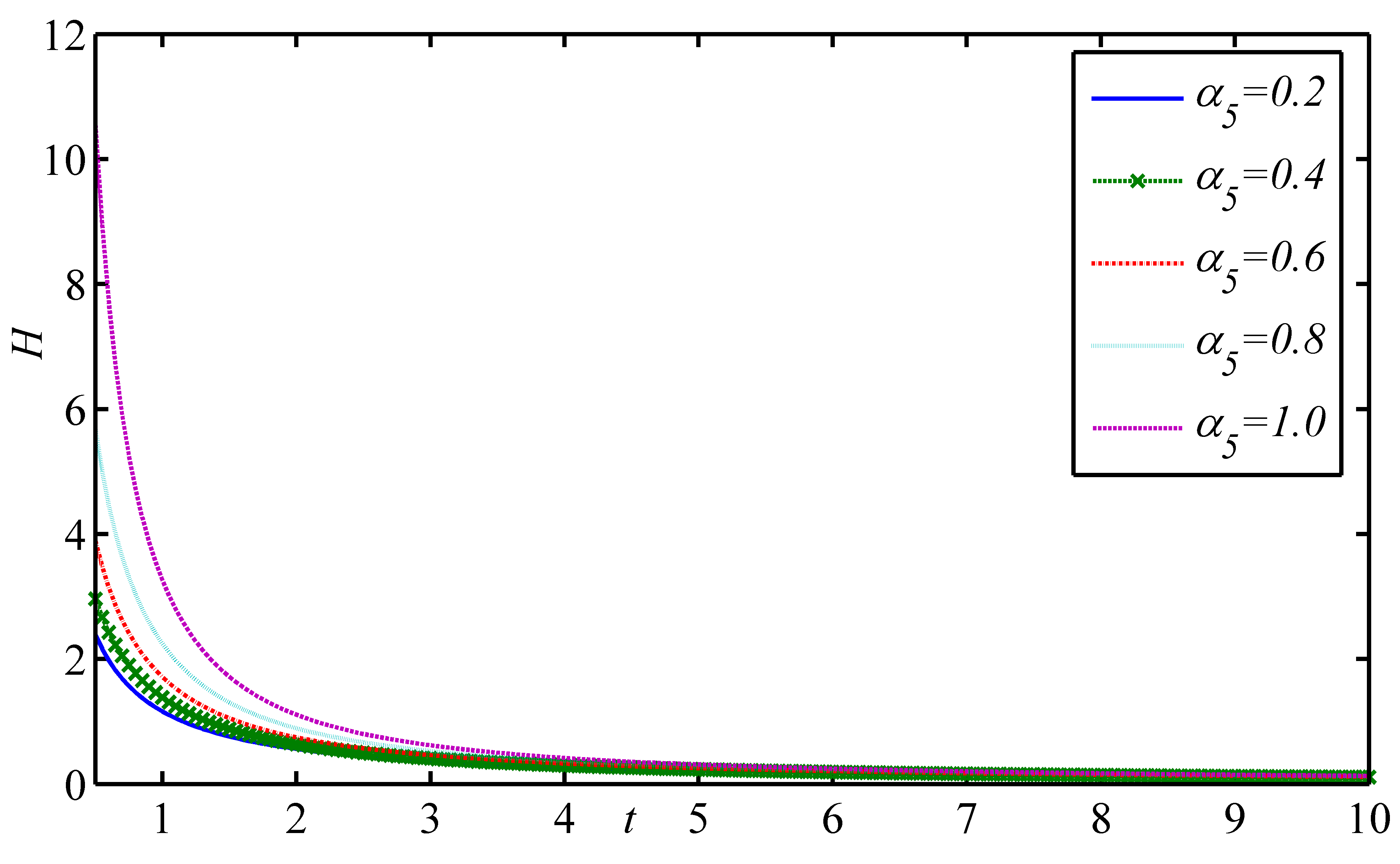}
        \caption{Variation of Hubble parameter against time  different $\alpha_5$}
        \label{fig19}
    \end{minipage}
\end{figure}

\begin{figure}[!htb]
  \centering
  \includegraphics[width=0.5\linewidth, height=0.25\textheight]{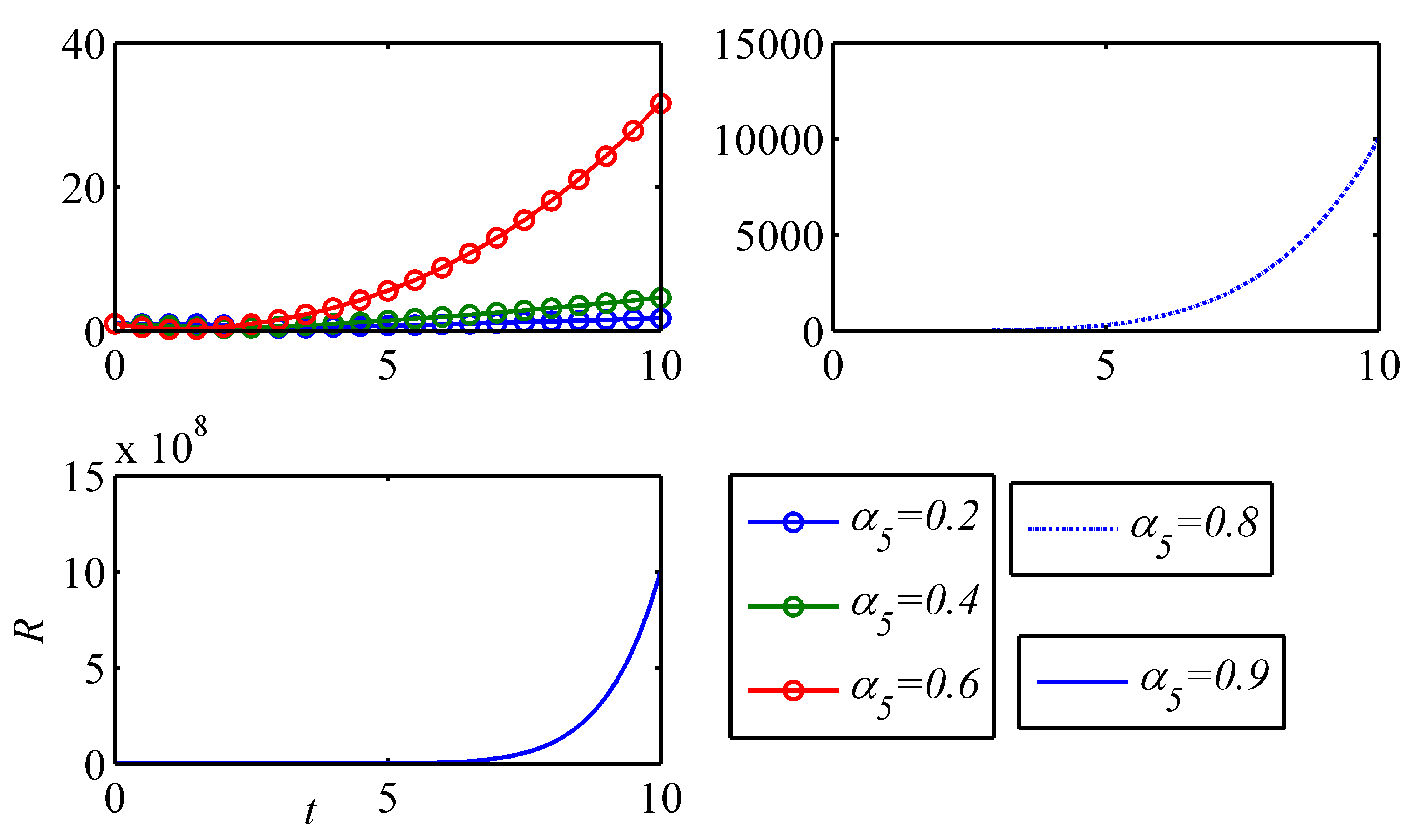}
  \caption{Variation of scale factor against time  for different $\alpha_5$}
  \label{fig20}
\end{figure}
\begin{figure}[!htb]
  \centering
  \includegraphics[width=0.5\linewidth, height=0.25\textheight]{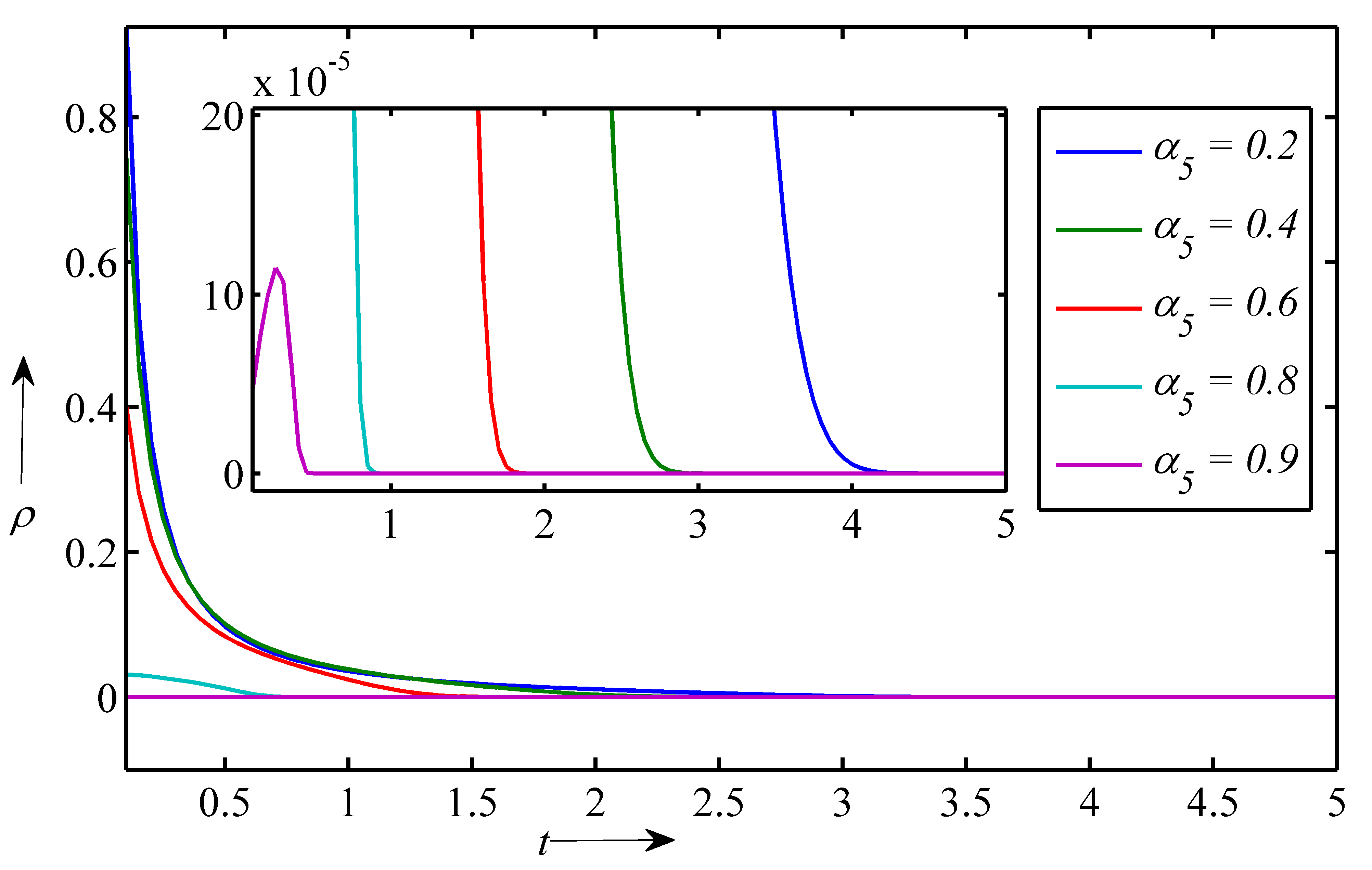}
  \caption{Variation of energy density $\rho$ against time  for $\alpha_1=0.5$, $\omega=1$, $\phi_0=1$, $c_2=0.1$ and different $\alpha_5$.}
  \label{fig21}
\end{figure}
\begin{figure}[!htb]
\begin{subfigure}
  \centering
  \includegraphics[width=.5\linewidth]{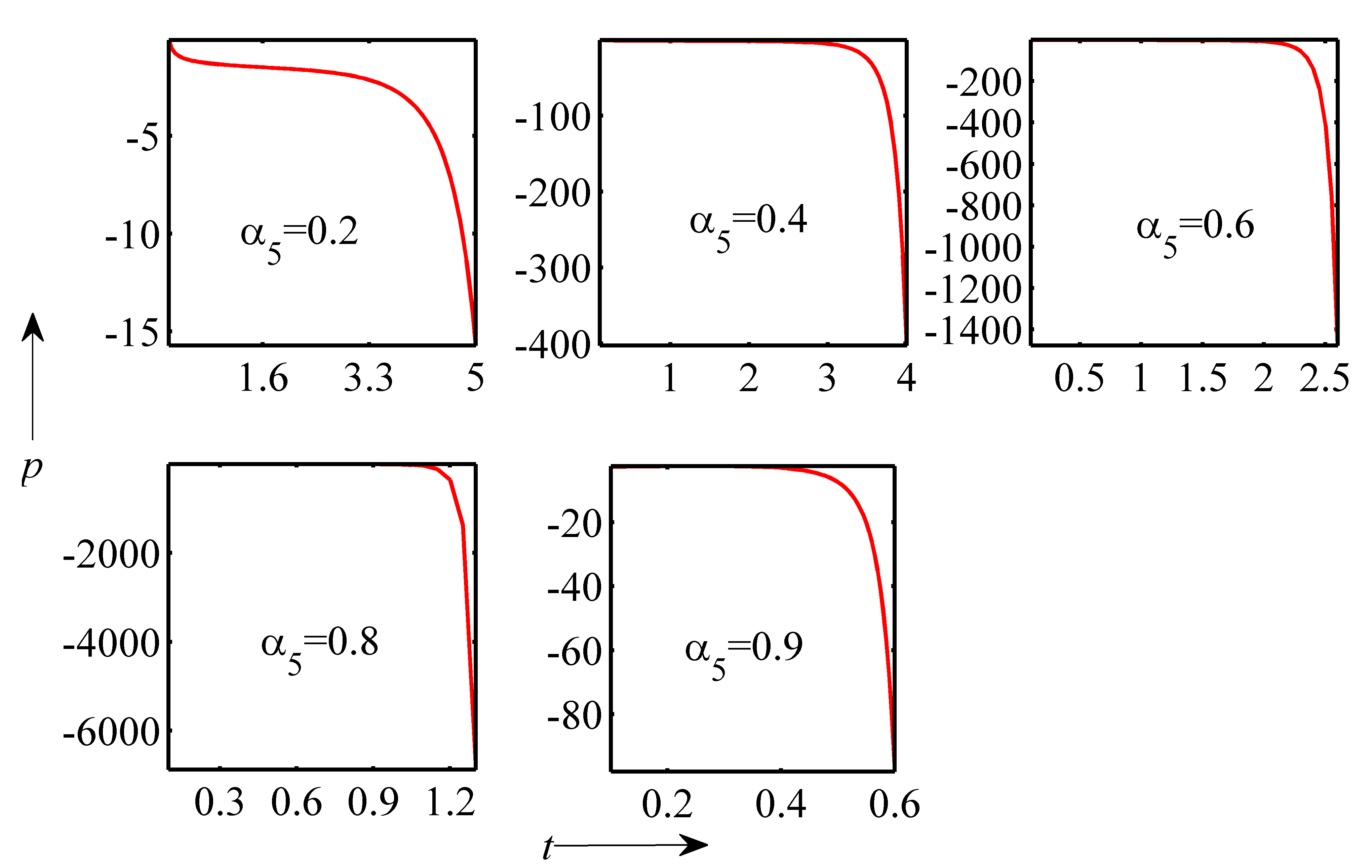}
\end{subfigure}%
\begin{subfigure}
  \centering
  \includegraphics[width=.5\linewidth]{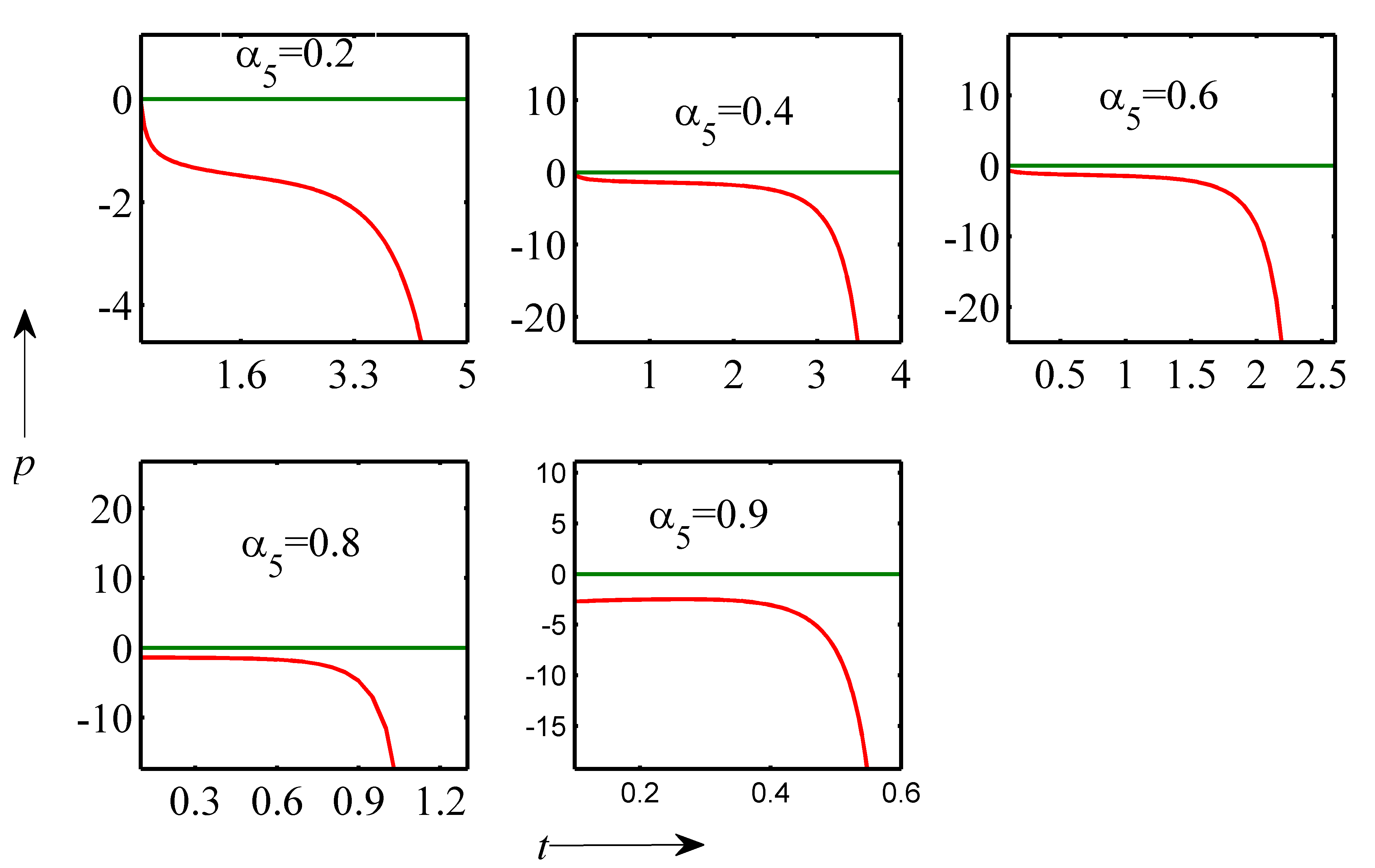}
\end{subfigure}
\caption{Variation of pressure $p$ against time  for $\alpha_1=0.5$, $A=1$, $B=1$, $n=0.1$, $\omega=1$, $\phi_0=1$, $c_2=0.1$ and different $\alpha_5$. Right panel  shows the zooming of the left panel  figures}
\label{fig22}
\end{figure}

The profile of deceleration parameter, Hubble parameter and scale factor against time is plotted in the Figure \ref{fig18}, Figure \ref{fig19} and Figure \ref{fig20} respectively for model-III. The observations are as follows:
\begin{itemize}
  \item Deceleration parameter $q$ is negative valued function of time and approaches towards zero with the evolution of time. In other words we can say, at early time our Universe is accelerating and follow an expansion with constant rate at late time (see Figure \ref{fig18}).
  \item Hubble parameter $H$ is a decreasing function of time and $H\rightarrow 0$ when $t\rightarrow\infty$. Also in this case higher the value of $\alpha_5$, higher is the value of Hubble parameter (see Figure \ref{fig19}).
  \item Scale factor $R$ is an increasing function of time and $R\rightarrow\infty$ when $t\rightarrow\infty$. Equation \eqref{eqn40} indicates that, $R$ is not defined for $\alpha_5=1$. As a representative case, we considered $0<\alpha_5<1$ (see Figure \ref{fig20}).
\end{itemize}
Figure \ref{fig21} and Figure \ref{fig22} depict the energy density and pressure profile against time respectively. For $\alpha_5\geq 1$, energy density possess physical unrealistic behavior, so $\alpha_5$ is restricted to $0<\alpha_5<1$. It is noticed that, energy density is a decreasing function of time  and $\rho\rightarrow 0$ when $t\rightarrow\infty$ (see Figure \ref{fig21}). Also pressure is a negative quantity with the evolution of time (see Figure \ref{fig22}).

\begin{figure}[!htb]
\begin{subfigure}
  \centering
  \includegraphics[width=.5\linewidth]{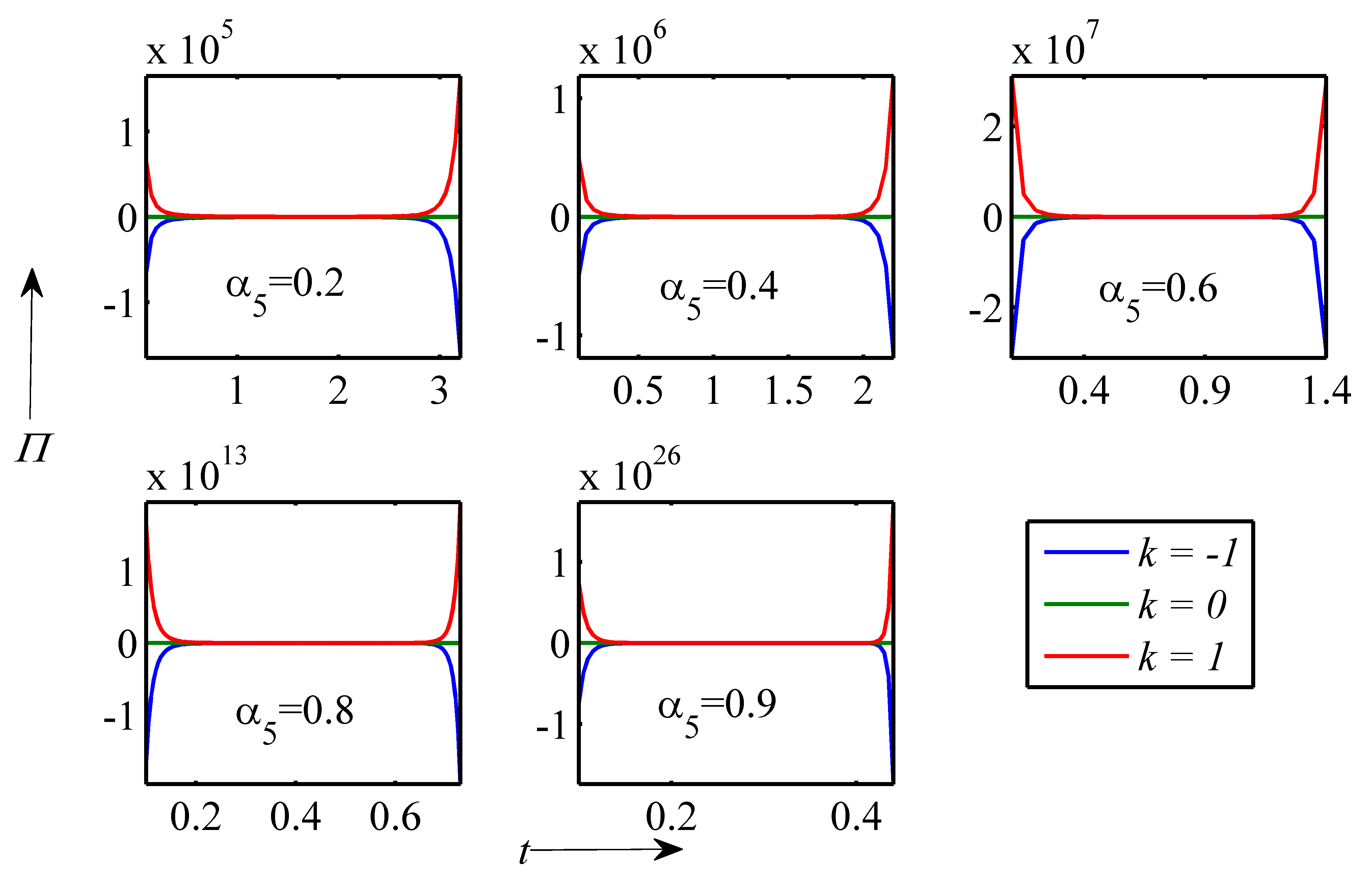}
\end{subfigure}%
\begin{subfigure}
  \centering
  \includegraphics[width=.5\linewidth]{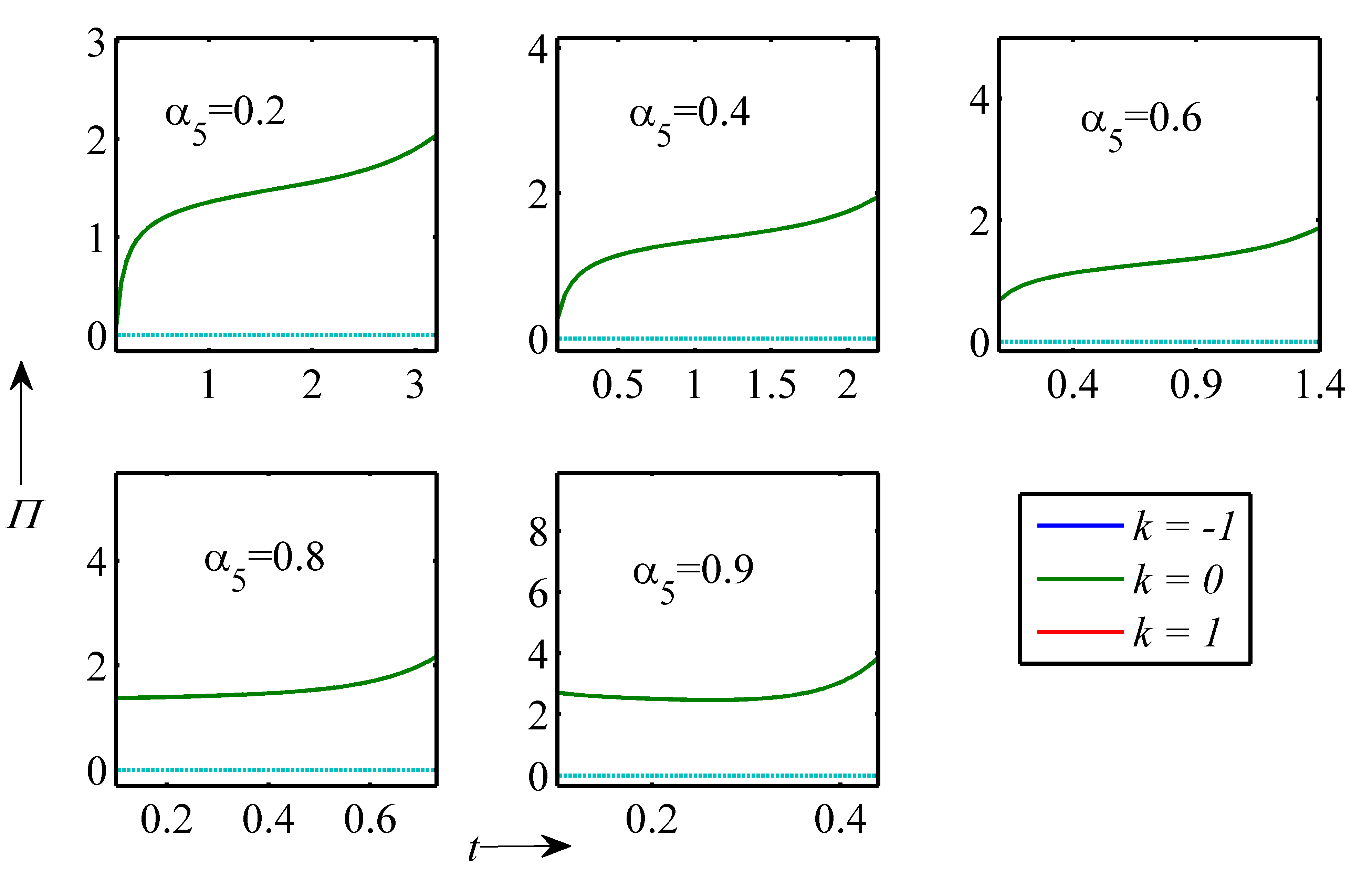}
\end{subfigure}
\caption{Variation of bulk viscous stress $\Pi$ against time  for $\alpha_1=0.5$, $A=1$, $B=1$, $n=0.1$, $\omega=1$, $\phi_0=1$, $c_2=0.1$ and different $\alpha_5$. Right panel  shows the zooming of the left panel  figures}
\label{fig23}
\end{figure}

\begin{figure}[!htb]
\begin{subfigure}
  \centering
  \includegraphics[width=.5\linewidth]{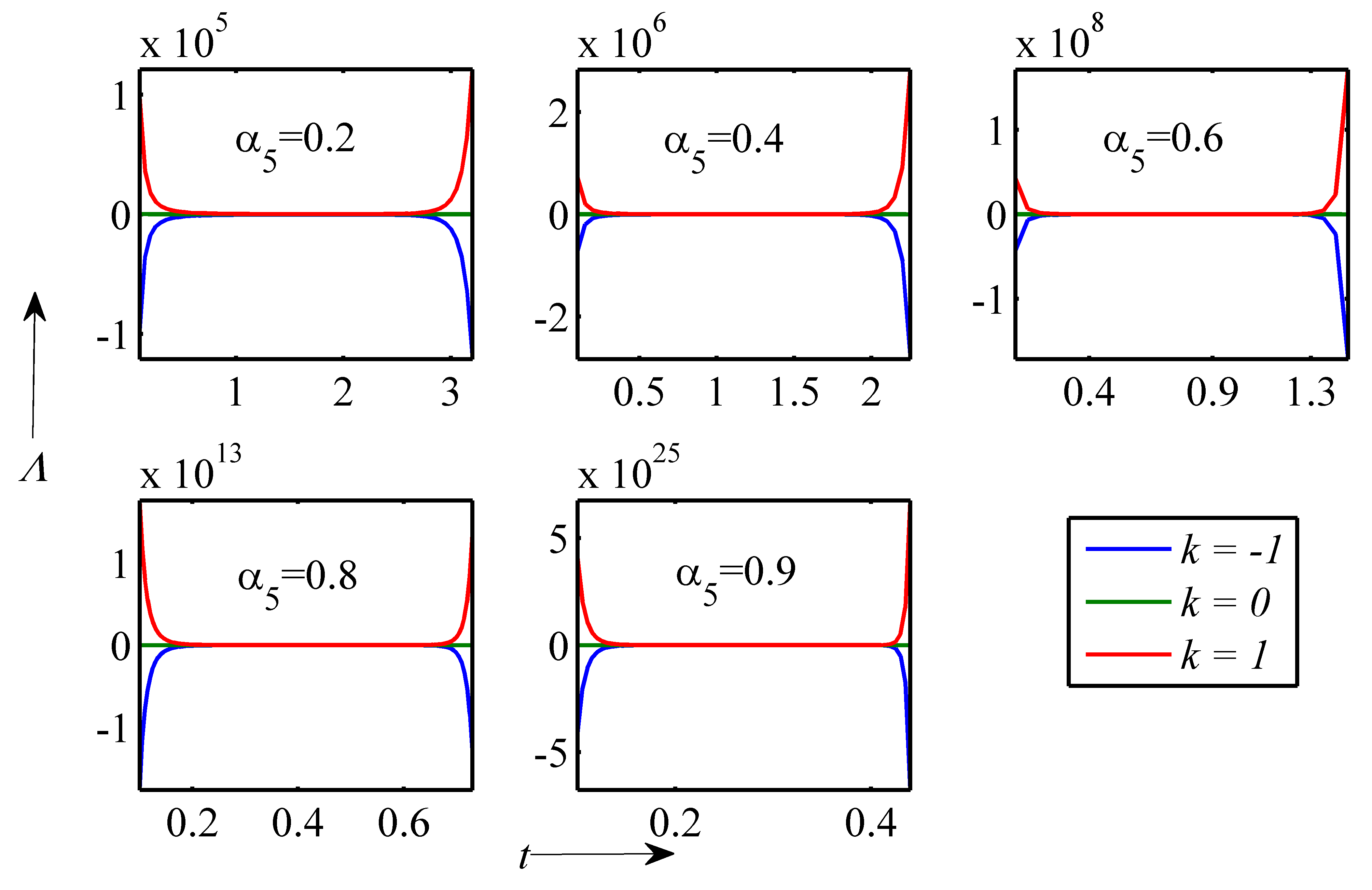}
\end{subfigure}%
\begin{subfigure}
  \centering
  \includegraphics[width=.5\linewidth]{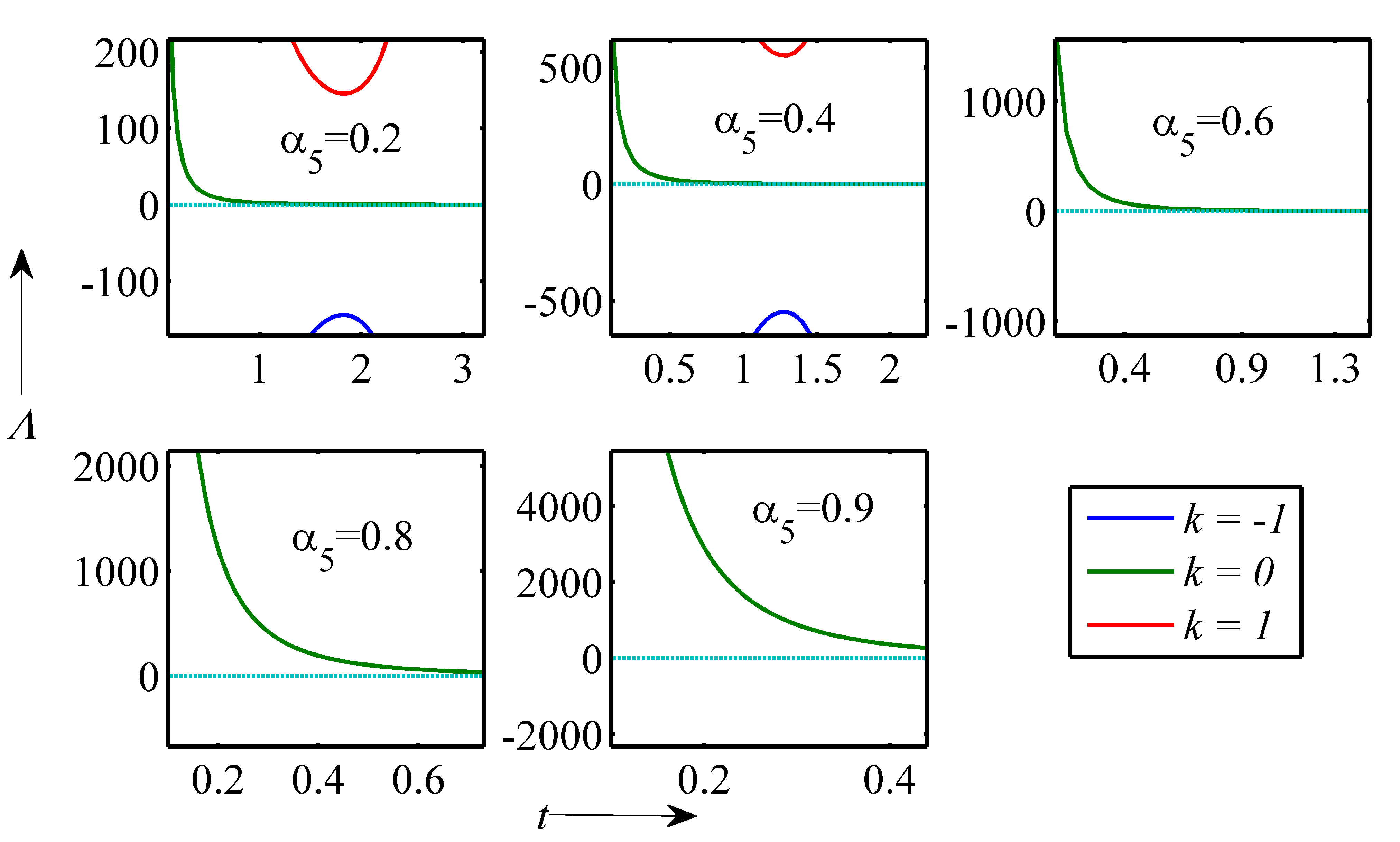}
\end{subfigure}
\caption{Variation of cosmological constant $\Lambda$  against time  for $\alpha_1=0.5$,  $\omega=1$,  $c_2=0.1$ and different $\alpha_5$. Right panel  shows the zooming of the left panel  figures}
\label{fig24}
\end{figure}

The profile of bulk viscous stress and cosmological constant against time is depicted in the Figure \ref{fig23} and Figure \ref{fig24} respectively for model-III. Bulk viscous stress is positive valued for flat and closed Universe whereas negative value for open Universe. Similar quantitative behaviour is observed for cosmological constant. In case of flat Universe, cosmological constant is a decreasing function of time and tending to zero with the evolution of time.

\section{Final statements}
In this article, we have studied the FRW cosmological model with modified Chaplygin gas in the framework of Brans-Dicke theory. The approximated exact solution is obtained for modified Einstein's field equation with the help of proposed form of deceleration parameter as in equation \eqref{eqn9}. We have presented three different cosmological models based on the choice of $\alpha_2$ and $\alpha_3$. The physical parameters involved in these three models are physically acceptable for some interval of $\alpha_2$ and $\alpha_3$, which follow the observational data. Here we would like conclude that, for physically acceptable cosmological models the choice of $\alpha_2$ and $\alpha_3$ are crucial.

\end{document}